\documentclass[
aps,%
11pt,%
final,%
notitlepage,%
oneside,%
onecolumn,%
nobibnotes,%
nofootinbib,%
superscriptaddress,%
noshowpacs,%
centertags]%
{revtex4}

\begin{document}

\title{GRAVITATIONAL LENSING IN PLASMIC MEDIUM}

\author{\firstname{G.~S.}~\surname{Bisnovatyi-Kogan}}
\email{gkogan@iki.rssi.ru}
\affiliation{Space Research Institute of Russian Academy of Sciences, Profsoyuznaya 84/32, Moscow 117997, Russia} \affiliation{National Research Nuclear University MEPhI (Moscow Engineering Physics Institute), Kashirskoe Shosse 31, Moscow 115409, Russia}
\author{\firstname{O.~Yu.}~\surname{Tsupko}}
\email{tsupko@iki.rssi.ru}
\affiliation{Space Research Institute of Russian Academy of Sciences, Profsoyuznaya 84/32, Moscow 117997, Russia} \affiliation{National Research Nuclear University MEPhI (Moscow Engineering Physics Institute), Kashirskoe Shosse 31, Moscow 115409, Russia}


\begin{abstract}
The influence of plasma on different effects of gravitational lensing is reviewed. Using the Hamiltonian
approach for geometrical optics in a medium in the presence of gravity, an exact formula for the photon
deflection angle by a black hole (or another body with a Schwarzschild metric) embedded in plasma with
a spherically symmetric density distribution is derived. The deflection angle in this case is determined by the
mutual combination of different factors: gravity, dispersion, and refraction. While the effects of deflection by
the gravity in vacuum and the refractive deflection in a nonhomogeneous medium are well known, the new
effect is that, in the case of a homogeneous plasma, in the absence of refractive deflection, the gravitational
deflection differs from the vacuum deflection and depends on the photon frequency. In the presence of a
plasma nonhomogeneity, the chromatic refractive deflection also occurs, so the presence of plasma always
makes gravitational lensing chromatic. In particular, the presence of plasma leads to different angular positions
of the same image if it is observed at different wavelengths. It is discussed in detail how to apply the presented
formulas for the calculation of the deflection angle in different situations. Gravitational lensing in
plasma beyond the weak deflection approximation is also considered.
\end{abstract}

\maketitle

\section{Introduction}

\begin{figure}
\centerline{\hbox{\includegraphics[width=0.7\textwidth]{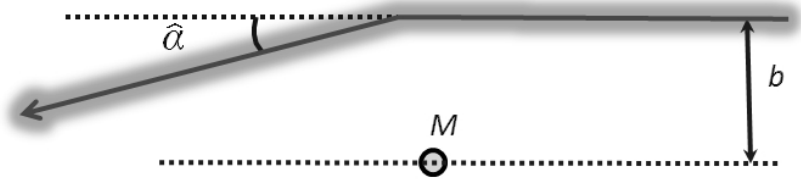}}} \caption{Gravitational deflection of the light ray by the massive body, in vacuum. The light ray passing near the spherical body of mass $M$ with impact parameter $b$ is deflected by the small angle $\hat{\alpha}$, see (\ref{einstein}). This picture corresponds situation when the impact parameter $b$ is much larger than the Schwarzschild radius of the gravitating body. Einstein deflection does not depend on the photon frequency.} \label{figure-einstein-angle}
\end{figure}

A gravitational lensing problem combines a well developed theory and a wide range of observational phenomena connected with the deflection of light rays by gravity. The gravitational lens theory usually deals with a geometrical optics in vacuum and uses a notion of the deflection angle. Basic assumption is an approximation of a weak deflection angle of a photon. General Relativity predicts that a light ray passing near a spherical body of mass $M$ with large impact parameter $b$ is deflected by the small angle (Fig.~\ref{figure-einstein-angle}):
\begin{equation} \label{einstein}
\hat{\alpha} = \frac{2 R_S}{b} = \frac{4M}{b}, \quad G=c=1 \, .
\end{equation}
This expression is valid if $b \gg R_S$, where $R_S = 2M$ is the Schwarzschild radius of the gravitating body. Deflection angle (\ref{einstein}) is usually called 'Einstein angle'. In the most astrophysical situations related with gravitational lensing the approximation of weak deflection is well satisfied. Trajectories of photons in vacuum, as well as deflection angles, do not depend on the photon frequency (or energy), so gravitational lensing is vacuum is achromatic.

The first gravitational lens outside the Solar system was discovered by Walsh et al \cite{Walsh1979}, as two images of the quasar 0957 + 561. The correlated variability of these images had confirmed that both images belong to the same object, and are formed due to light deflection by the gravitation of a massive intermittent object. Tens of gravitational lens examples have been discovered in subsequent years, see, for example, \cite{Castles}.

In this paper we review an influence of plasma on different effects of gravitational lensing \cite{BKTs2009}, \cite{BKTs2010}, \cite{TsBK2012}, \cite{TsBK2013}. The principal feature is that the deflection angles are changed in presence of plasma, and chromatic deflection occurs. In particular, it leads to angular difference in positions of images at different wave frequencies. For the light rays propagating through cosmic plasma such effects may be significant only for very long radiowaves.\footnote{In vacuum  the chromatic effects of gravitational lensing take place in study of microlensing of quasars by individual stars in lens (galaxy). Physical sources like quasars may have different sizes for different wavelengths. Magnification and the light curves expected as a result of microlensing depend on the source size, so chromatic effects arise. Such chromatic effects can give information about source structure, see \cite{Kayser1986}, \cite{WambsganssPaczynski1991}. Sometimes the term 'chromatic gravitational lensing' is used for this situation. Rigorously speaking, this situation is 'achromatic lensing of chromatic source' rather than 'chromatic gravitational lensing'. In this paper we consider chromatic effects which related with the situation when the photon deflection itself becomes chromatic, in particular, the gravitational deflection of photon in plasma is chromatic.}

There exists another possible reason of curving of light trajectory: refractive deflection. In non-homogeneous transparent media the photon moves along the curved trajectory, it is called as refraction. Refraction is widely known phenomenon, and it is easily observed at boundaries between water, air or glass, which is described by Snell's law. Working in the frame of geometrical optics, we can characterize the properties of the medium by its refractive index. In inhomogeneous medium the refractive index depends explicitly on the space coordinates, and shape of trajectory is determined by this dependence. The refractive deflection of light has no relation to relativity and gravity, and takes place only in non-homogeneous media. If medium is dispersive too, which means that the refractive index depends on the frequency, then the refractive deflection is also chromatic.

Considering propagation of a light under action of both gravity and plasma non-homogeneity, in linear approximation we can just summarize two deflection angles: a vacuum deflection due to gravitation of a point mass, and a refractive deflection due to the non-homogeneity of the plasma. These effects are separated from each other. This approach was used in consideration of deflection of radio rays passing the Sun \cite{Muhleman1970} where the deflection was a separable combination of general relativistic deflection in vacuum and refraction in the coronal electron plasma, see also \cite{Muhleman1966}, \cite{zadachnik}. The same approach is used in book of Bliokh and Minakov \cite{BliokhMinakov} where gravitational
lensing was considered, produced by a gravitating body surrounded by a plasma with a spherically symmetric distribution.

The approach above, considering propagation of light in presence of gravity and medium as a simple sum of two deflections, does not contain nonlinear effects. Possible mutual influence of medium and gravity is not taken into account. To consider this problem more rigourously, we need theory which includes gravity and medium presence simultaneously.

A self-consistent approach to the light propagation in the gravitational field, in presence of a medium, was developed in a classical book of Synge \cite{Synge}. For discussion and application of the Synge's general relativistic Hamiltonian theory for the geometrical optics see also papers of Bi\v{c}\'{a}k and Hadrava \cite{Bicak}, Kichenassamy and Krikorian \cite{Krikorian1985}, Krikorian \cite{Krikorian1999}.

Comprehensive review of general-relativistic ray optics in media is presented in the monograph of Perlick \cite{Perlick2000}. He used a different type of the approach to the problem. As one of applications of his formalism, general formulae for the light deflection angle in the Schwarzschild and Kerr metrics, in presence of spherically distributed plasma, have been obtained in the integral form. Formula for the Schwarzschild metric was later rederived on basis of Synge's approach in the paper \cite{TsBK2013}.

Investigations of non-linear effects connected with combined action of gravity and plasma were performed in papers of Bisnovatyi-Kogan and Tsupko \cite{BKTs2009}, \cite{BKTs2010}, \cite{TsBK2012}, \cite{TsBK2013}, on the basis of Synge's theory. The main idea was to investigate the influence of plasma on the light trajectory in a homogeneous plasma where there is no  refractive deflection. It was shown that the gravitational deflection itself in the medium is not the same as in vacuum. Therefore, if we would like to consider tiny effects we should use corrected formula for the gravitational deflection instead of the vacuum gravitational deflection.

This idea is the clearest in case when plasma is homogeneous. In this case the refractive deflection is absent. The question can arise: is the gravitational deflection of the light rays in this case the same as in vacuum? The answer is the following: the deflection will be differ from vacuum case \cite{BKTs2009}. For static plasma  with the refractive index
\begin{equation}
n^2 = 1 - \frac{\omega_e^2}{[\omega(r)]^2} \, , \quad \omega_e^2 = \frac{4 \pi e^2 N_0}{m} = \mbox{const}
\end{equation}
in a static weak gravitational field of a point mass, the gravitational deflection angle will be described by the simple formula \cite{BKTs2009}:
\begin{equation} \label{main-res}
\hat{\alpha} = \frac{R_S}{b} \left( 1 + \frac{1}{1 - (\omega_e^2 / \omega^2)} \right).
\end{equation}
Here  the frequency of the photon $\omega(r)$ depends on the space coordinate $r$ due to the presence of a gravitational field (gravitational redshift). We denote $\omega(\infty) \equiv \omega$, $e$ is the charge of the electron, $m$ is the electron mass, $\omega_e$ is the electron plasma frequency, $N_0 = \mbox{const}$ is the electron concentration of a homogeneous plasma. This formula is written for homogeneous plasma and valid under the condition of smallness of $\hat{\alpha}$ ($b \gg R_S$). At $\omega_e=0$ (concentration $N_0=0$) or $\omega \rightarrow \infty$ this formula turns into the deflection angle (\ref{einstein}) equal to $2R_S/b$.

Thus, it was shown for the first time in \cite{BKTs2009}, \cite{BKTs2010} that due to dispersive properties of plasma even in the homogeneous plasma the gravitational deflection differs from vacuum deflection angle, and gravitational deflection angle in plasma depends on frequency of the photon. By using Synge's approach, it was also found \cite{BKTs2009}, \cite{BKTs2010} that the photon deflection is the same as in vacuum, if medium is homogeneous and non-dispersive (refraction index $n$=const). The physical reason is a dependence
of the wave frequency on space coordinates in presence of gravity (gravitational redshift). In the paper \cite{BKTs2010} we have considered the gravitational lensing of radiowaves by the point or the spherical body in presence of a homogeneous and non-homogeneous plasma, together with the refractive deflection due to the plasma inhomogeneity, and comparison of these two effects have been there performed. All these results have been obtained in the approximation of smallness of $\hat{\alpha}$.

Er and Mao \cite{Mao2014} have studied the gravitational lensing when non-homogeneous plasma surrounds the lens. Using formulas for the deflection from \cite{BKTs2009}, \cite{BKTs2010}, they have calculated the extra deflection angle induced by the plasma for non-homogeneous plasma models. They have performed  numerical modelling of different effects caused by this extra deflection. In particular they have shown that a plasma presence may cause significant effects to be detected in low frequency radio observations (a few hundred MHz): the change of position due to a plasma can reach a few tens of milli-arcsec.

Recently Morozova et al \cite{morozova} have considered a gravitational lensing by a rotating massive object in a plasma. Their approach is based on papers \cite{BKTs2009}, \cite{BKTs2010}, they additionally take into account the rotation of lens and investigate new effects connected with presence of rotation.

\begin{figure}
\centerline{\hbox{\includegraphics[width=0.40\textwidth]{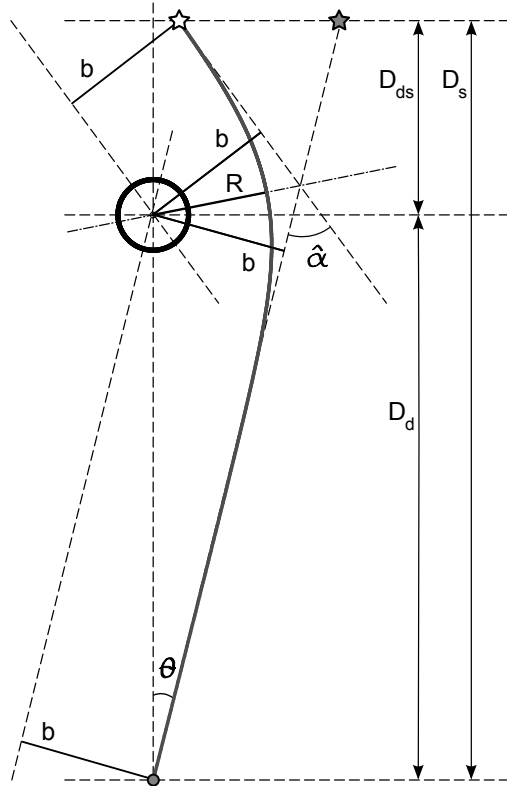}}} \caption{Change of apparent angular position of point source due to gravitational lensing. Trajectory of light ray is calculated in the Schwarzschild metric. Light ray from the source deflected by angle $\hat{\alpha}$ by the point-mass gravitational lens with the Schwarzschild radius $R_S$ goes to the observer. The observer sees the image of the source at angular position $\theta$, which is different from the source position. $R$ is the closest point of trajectory to gravitating center, it is usually referred as the distance of the closest approach, $b$ is the impact parameter of the photon, $D_d$ is the distance between observer and lens, $D_s$ is the distance between observer and source, $D_{ds}$ is the distance between lens and source. We should comment that at this picture the deflection angle is not small actually, and can not be calculated by approximate formula (\ref{einstein}), we draw situation with such angle for visualization.} \label{figure-prd1}
\end{figure}

\begin{figure}
\centerline{\hbox{\includegraphics[width=0.4\textwidth]{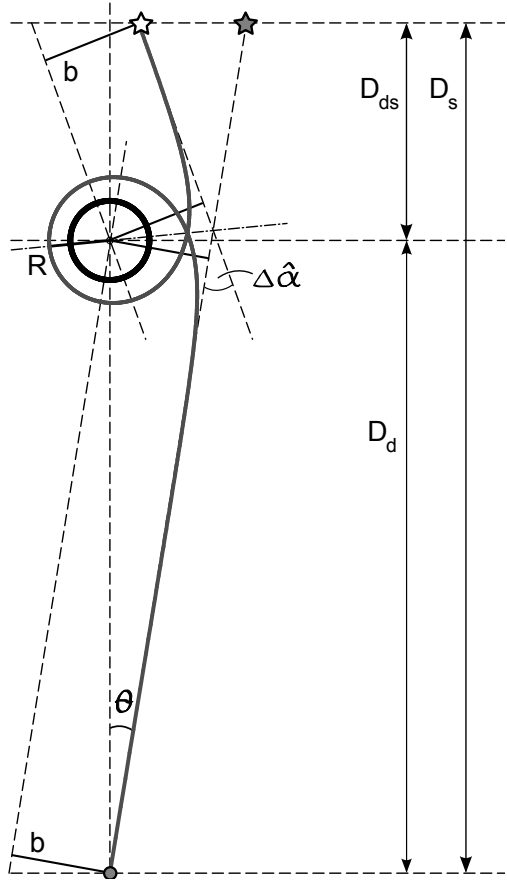}}} \caption{A scheme of formation of the relativistic image of the point source. Trajectory of light ray is calculated in the Schwarzschild metric. Before reaching observer, light ray perform turn around of black hole, at radius close to $r_M = 3R_S/2$. Deflection angle of the photon in this case equals to $\hat{\alpha} = 2\pi + \Delta \hat{\alpha}$. There exist rays which make more revolutions around of black hole and form other relativistic images, the closest approach distances $R$ for these rays are closer to $r_M$. Combining with relativistic images at another side from the lens, these images form two infinite sequences of images.} \label{figure-prd2}
\end{figure}

Theory of gravitational lensing usually considers small deflection angles ($\hat{\alpha} \ll 1$), see Fig.~\ref{figure-prd1}. Light rays which are deflected by angles about $2\pi$, $4\pi$, ... can also form images. If the photon impact parameter is close to its critical value, a photon which goes from infinity can perform several turns around the central object and then go to infinity. Such photons form images (Fig.~\ref{figure-prd2}), which are called relativistic images \cite{Virbhadra2000}. Influence of plasma on properties of relativistic images was studied for the first time in \cite{TsBK2013}.

The paper is organized as follows. In Section 2 we briefly describe the basic effects of gravitational lensing. In section 3 we consider the gravitational lensing in the case of a strong deflection of  photons in vacuum. In section 4 we derive an exact formula for the deflection angle of photons in the Schwarzschild metric  with spherically symmetric distribution of plasma. In section 5 we derive the exact formula for the angle of  gravitational deflection of light in a homogeneous plasma. In section 6 we consider the gravitational deflection in the homogeneous plasma in the case of a weak deflection. In section 7 we consider weak deflection of the photon in presence of gravity and non-homogeneous plasma. In section 8 we consider the gravitational deflection of light in the homogeneous plasma in strong deflection limit and discuss influence of plasma on relativistic images. In section 9 we make concluding remarks.

\begin{figure}
\centerline{\hbox{\includegraphics[width=0.6\textwidth]{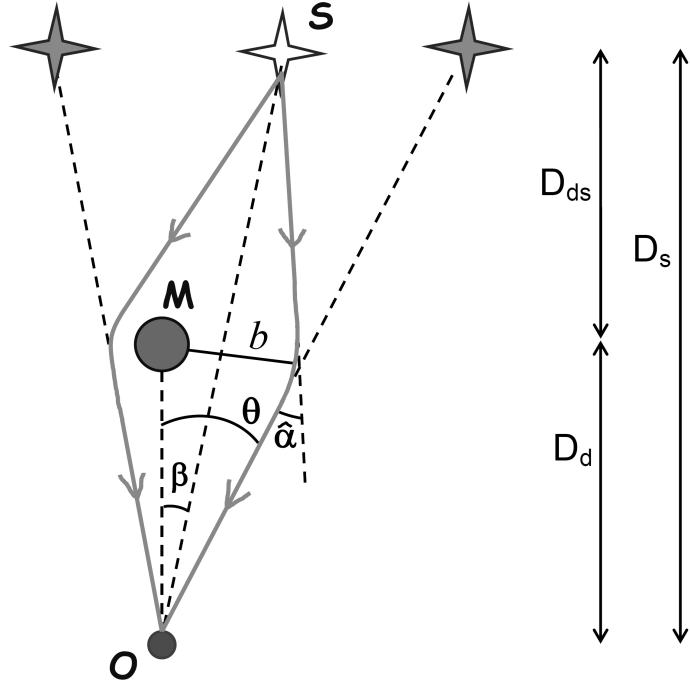}}} \caption{Lensing of the point source by the Schwarzschild point-mass lens in vacuum. Light rays from the source $S$ are deflected by angle $\hat{\alpha}$ by the point mass $M$. Observer $O$ sees two images at opposite sides relatively to lens, instead of the real source $S$. The image which situated at the same side from the lens as real source is called primary image, another one is secondary. Rays from source which are not in plane of Source-Lens-Observer will not go to observer, therefore in case of point source the observer sees two images as point images.} \label{figure-schwlens}
\end{figure}

\begin{figure}
\centerline{\hbox{\includegraphics[width=0.6\textwidth]{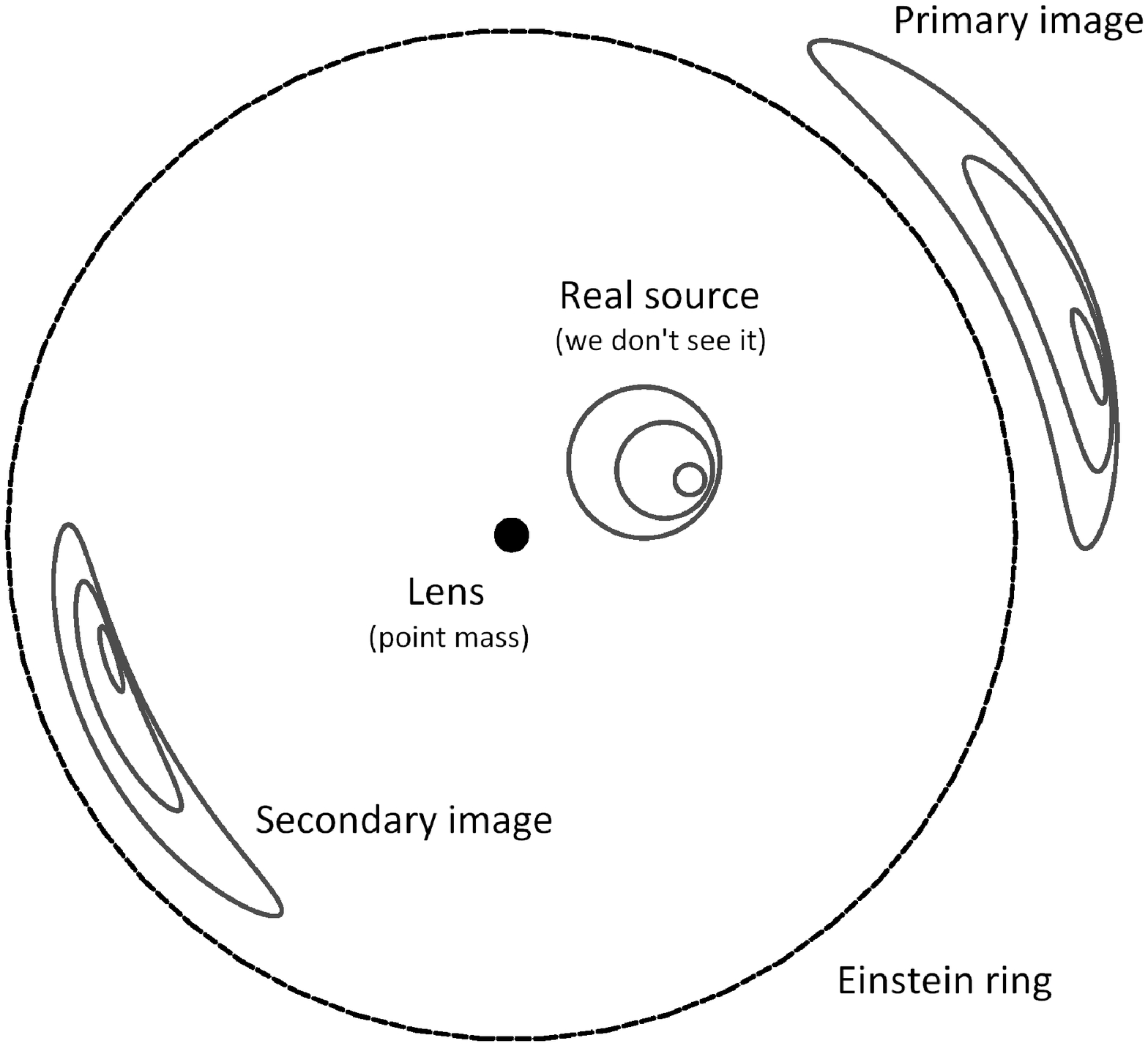}}} \caption{
Primary and secondary images in case of lensing by the Schwarzschild point-mass lens in vacuum. Source is extended (not point but circle) and has internal structure drawn by curves inside circle, see picture. Observer does not see real source and sees two prolongated images with another angular sizes. At this picture the angular size of Einstein ring is also presented.} \label{figure-primsec}
\end{figure}

\begin{figure}
\centerline{\hbox{\includegraphics[width=0.7\textwidth]{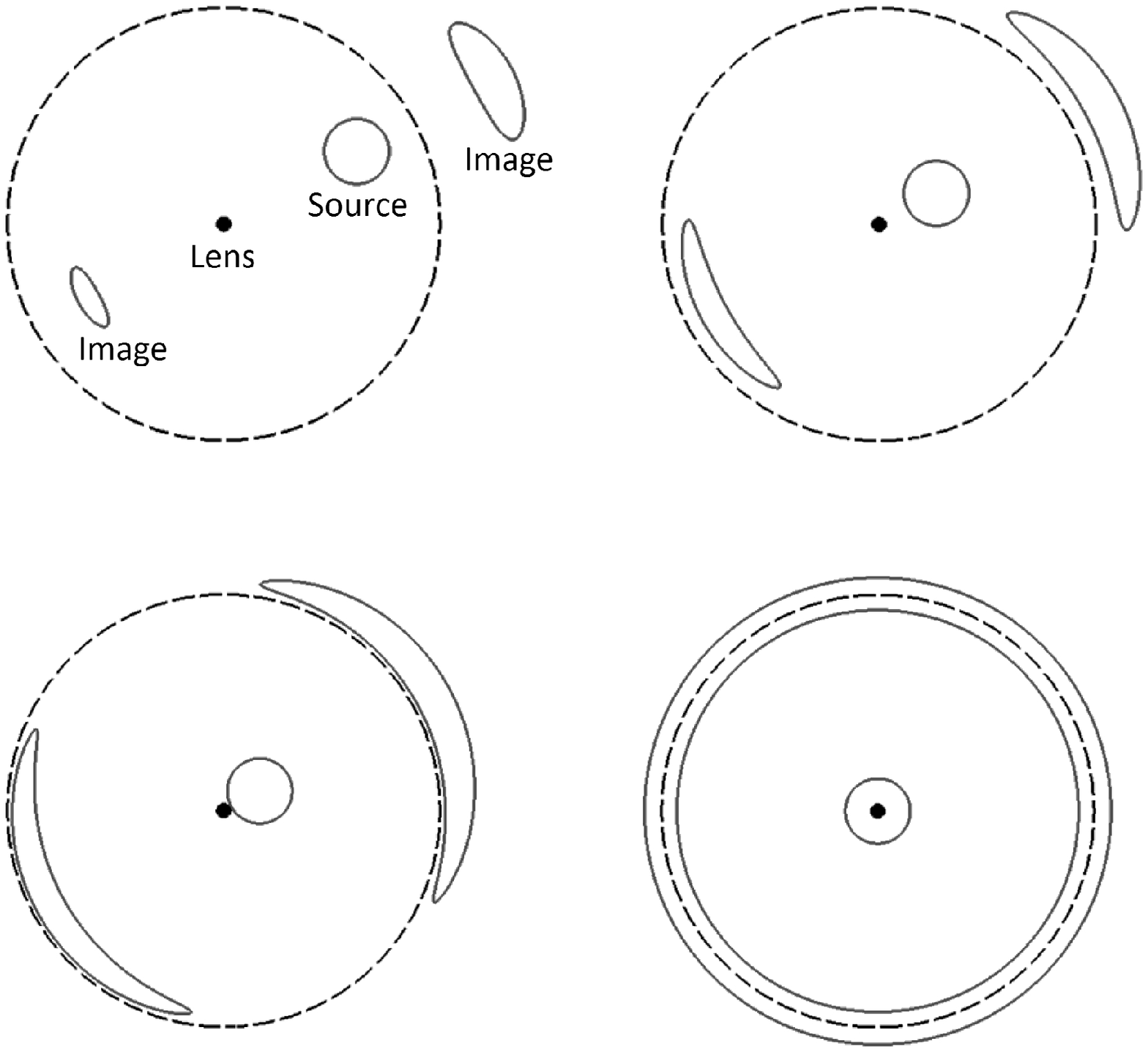}}} \caption{
Formation of arcs in case of lensing by the Schwarzschild point-mass lens in vacuum. Different angular positions of source relatively to lens are presented. Primary and secondary images become more prolongated if source is closer to lens (closer to line passing through observer and lens). In case of perfect alignment the observer sees ring.} \label{figure-arcs}
\end{figure}

\section{Basic effects of gravitational lensing}

In this section we make a short review of basic effects of gravitational lensing. For details, see, for example, books and reviews \cite{GL} - \cite{Hoekstra-review}.	

Gravitational lensing changes an apparent angular position of the source. Example of calculation of the trajectory of the photon for lensing with the small deflection is presented on Fig.~\ref{figure-prd1}. In this example the lens is a point body. Light ray from the source deflected by angle $\hat{\alpha}$ by the point-mass gravitational lens with the Schwarzschild radius $R_S$ goes to the observer. The observer sees the image of the source at angular position $\theta$ which is different from the source position. Bending of light from the distant stars by gravity of the Sun leads to apparent displacement of the stars position. It can be observed for stars near Sun disc during eclipse, and it was one of the first observational proof of General Relativity \cite{Dyson1920}.

Gravitational lensing leads to formation of multiple images, lensing on one gravitating point mass leads, in general, to formation of two images. On Fig.~\ref{figure-schwlens} there is the example of the simplest model of lens, the Schwarzschild point-mass lens. Due to bending of light the observer sees two images of the same source at different sides from lens. If source is point-like, then images of source are two points. If the source has a finite size, the deformation of images occurs, which is shown on Fig.~\ref{figure-primsec} and Fig.~\ref{figure-arcs} for source with circular shape. If source, lens and observer are aligned enough, the deformation can be rather big, and observer sees two arcs. Deformation becomes bigger, if angular position of source becomes closer to angular position of lens. In the case of perfect alignment, observer sees image in the form of ring --- Einstein (Einstein-Chwolson) ring. If the gravitational lens is not point-like, and has more complicated mass distribution, observer can see more than two images of the same source. In modelling of real system with multiple images the point mass lens is never used. Observed gravitational lens system are described usually by lens with broken spherical symmetry.

Another effect of gravitational lensing is a so called magnification effect \cite{GL}. The surface brightness for the image is identical to that of the source in the absence of the lens. The flux of the image  is the product of its surface brightness and the solid angle it subtends on the sky. The flux from source is changed due to gravitational lensing, because angular size of source is changed. Angular size (and flux) of each image is different from a source, and can be larger and smaller. The magnification factor $\mu$ is the ratio of the flux of the image to the flux of the unlensed source, and is equal to the ratio of the solid angles of the image, and the unlensed source: $\mu = \Delta \omega/(\Delta \omega)_0$.

Gravitaional lensing leads also to time delay: in presence of lens the time of light travelling from source to observer is changed \cite{GL}. Time delay consists of two parts: geometrical delay, due to changing of trajectory shape, and potential or Shapiro delay, due to propagation of light in gravitational field. In the case when multiple images are observed, every image has its own delay, and time delay between images can be found. It is interesting as a way of finding of Hubble constant, see \cite{Refsdal1964b}, \cite{GL2}.

From the observational point of view there are three types of the gravitational lensing.

(i) Strong lensing. In this case effect of lensing is strong what means the multiple imaging or strong deformation of image. Multiple images are close to each other at sky and have the same redshifts, what, together with spectral and time variation properties of the images, permits to obtain a clear conclusion about the presence of the gravitational lens. Extended arcs and rings are also examples of strong lensing.

(ii) Weak lensing. In this case the lensing images are characterized by weak distortions and small magnifications. Usually only one image of each source is visible. In this case we can not identify fact of lensing using individual source because we do not know a real position of the source, its shape, and its real luminosity. To conclude the presence of lensing in this case is possible only in a statistical sense. Observations of many weakly distorted images of galaxies give a possibility to reconstruct the mass distribution in the gravitational lens. The example of the reconstruction of the distribution of the dark matter gravitational potential in merging Bullet cluster of galaxies 1E0657- 558 ($z$=0.296)  is given in \cite{Clowe2006}. It was obtained due to observations by HST/ASC  of the field galaxies behind the clusters. This distribution is obtained together with the distribution of ordinary matter, by Chandra X-ray observations of the shocks formed in the merging cluster 1E0657-558, which brightness maxima  do not coincide with the maxima of the dark matter gravitational potential. These observations are considered as a direct empirical proof of the existence of a dark matter.

(iii) Microlensing. It is version of strong lensing in which the image separation is too small to be resolved. Lens which is a single star of about solar mass can lead to images split into microimages separated by microarcseconds (it explains an origin of the term). Multiple images due to stellar-mass objects can therefore not be resolved. The only observable effect of microlensing is the alteration of the apparent luminosity of background sources (change of flux), see \cite{Refsdal1964a}, \cite{Byalko1969}, \cite{GL}. There are several applications of microlensing effect. In the case of a quasar microlensing, multiple images (which occur due to strong lensing) of the quasar are studied. The images of a multiply imaged quasar are seen through a galaxy. Since galaxies contain stars, stellar-mass objects can affect the brightness of these images, and this is effect of microlensing for quasars. One of the most interesting application of microlensing is a microlensing by planets, which started with paper of Mao and Paczy\'{n}ski \cite{Mao1991}, for observation discovery see, for example \cite{Beaulieu2006}.

From physical point of view we can distinguish situations of lensing on basis of magnitude of deflection angle: weak deflection of photon (small deflection angles, $\hat{\alpha} \ll 1$) and strong deflection of photon (deflection angles are not small, $\hat{\alpha} \ge 1$).

In  most problems of the gravitational lensing the weak deflection approximation is applied. It means that the formula of Einstein deflection, in the case of a point lens, and formulae based on it, in the case of a mass distribution, are used. The lens equation \cite{GL} connects positions of source and images, at given geometry, in flat or expanded space-time. With a given mass distribution in the lens, the lens equation allows us to find positions, forms and magnifications of images, analytically or numerically.

\begin{figure}
\centerline{\hbox{\includegraphics[width=0.4\textwidth]{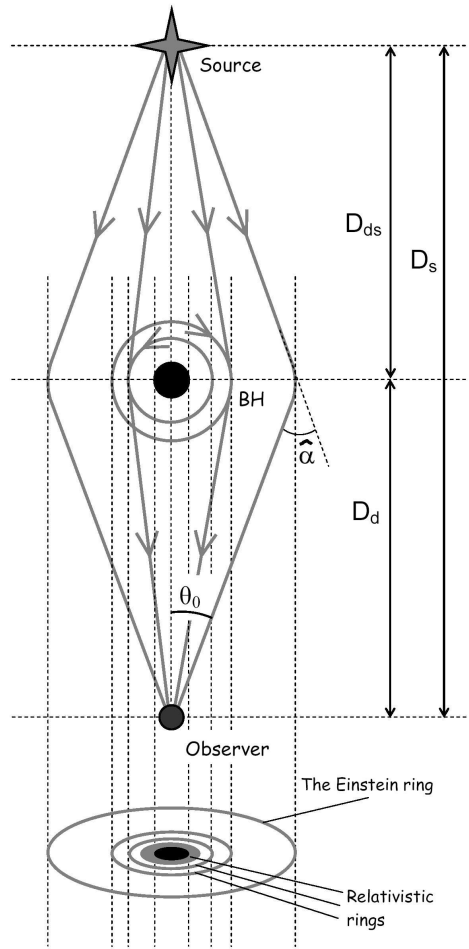}}} \caption{Einstein ring, and relativistic rings. If the source, the lens, and the observer are perfectly alighned, a circle, known as the Einstein ring, is formed. Inside this ring there are rings formed by photons which have been deflected by angles about 2$\pi$, 4$\pi$, 6$\pi$ ...; these rings are sometimes referred to as relativistic rings.} \label{figure-rings}
\end{figure}

\section{Gravitational lensing in vacuum in the case of a strong deflection}

Weak deflection approximation is enough for the most problems and applications of gravitational lensing. The example of situation, when the deflection angles of photons beyond usual weak deflection approximation are considered, is a lensing by black hole. In this case the photon can move close enough to central object to undergo large deflection.

Let us consider the motion of a photon in the neighborhood of a
black hole with a Schwarzschild metric. The
Schwarzschild metric is given by
\begin{equation}
ds^2 = - \left(1-\frac{2M}{r}\right)dt^2 + \frac{dr^2}{1-2M/r} \, + \, r^2(d\theta^2 + \sin^2 \theta \, d\varphi^2), \; R_S = 2 M, \; G=c=1.
\end{equation}

The shape of the orbit of a photon incident from infinity on a
black hole is determined by its impact parameter $b$ \cite{MTW}:

(i) if $b < 3\sqrt{3} M$, then the photon falls to $R_S = 2 M$ and
is absorbed by the black hole;

(ii) if $b > 3\sqrt{3} M$, then the photon is deflected by an angle
$\hat{\alpha}$ and flies off to infinity. In gravitational lensing we are interested in this case.

Exact expression for the deflection angle $\hat{\alpha}$ in the case of a motion in the Schwarzschild metric can be derived from equations
determining the photon orbit, see, for example, \cite{MTW, BKTs2008}. For a given mass $M$, the deflection angle of a photon is a function of the radius of the closest approach $R$, and is represented by the integral
\begin{equation}
\label{vacuum-exact} \hat{\alpha} = 2 \int \limits_R^\infty \frac{dr}{r^2 \sqrt{\frac{1}{b^2} - \frac{1}{r^2} \left( 1- \frac{2M}{r}
\right) }} - \pi \, .
\end{equation}
The impact parameter $b$, corresponding to the distance of the closest approach $R$ is written as
\begin{equation}
\label{b-and-R} b^2 = \frac{R^3}{R-2M}.
\end{equation}
The expression of the integral (\ref{vacuum-exact}) in terms of elliptical integrals was first obtained in \cite{Darwin1959}.
The expansions of exact deflection angle in powers of $M/R$ and $M/b$ is given in \cite{Keeton}.

In the case of large impact parameters $b \gg 3\sqrt{3} M$ the
orbit is almost a straight line with a small deflection by an
angle $\hat{\alpha} = 4M/b$ (see Fig.\ref{figure-einstein-angle}). In this case (weak deflection limit) the
impact parameter and the distance of closest approach are almost
the same (see Fig.\ref{figure-prd1}). We can neglect a difference between the impact parameter
and the distance of closest approach, and write formula (\ref{einstein}) either with $b$ or $R$. A form of the
formula (\ref{einstein}) in textbooks depends usually on a way of derivation (compare, for example, \cite{LL2} and \cite{MTW}). This case (weak deflection limit) is considered usually in the theory of gravitational lensing.

There is another limiting case in which the photon deflection angle can be written in the analytical form. It is the case when the deflection angle is large, it is usually referred as strong deflection limit. If the value of the impact parameter is close to the critical value, $0 < b/M - 3\sqrt{3} \ll 1$, then the photon makes one or
several turns around the black hole near a radius $r_M = 3M$ and
flies off to infinity (Fig.\ref{figure-prd2}). In this case (the strong deflection limit) the deflection angle is \cite{Darwin1959},
\cite{Bozza2001}
\begin{eqnarray}
\label{vacuum-alpha-R}
\hat{\alpha} &=& - 2 \ln \left( \frac{R}{r_M} - 1\right) +  2 \ln[12(2-\sqrt{3})] - \pi = \nonumber\\
&=& - 2 \ln \frac{R-3M}{36(2-\sqrt{3}) M} - \pi = - 2 \ln \frac{(R-3M)(2+\sqrt{3})}{36M} - \pi \, ,
\end{eqnarray}
or, as a function of the impact parameter $b$ \cite{Bozza2001}, \cite{Bozza2002}
\begin{eqnarray}
\label{vacuum-alpha-b}  \hat{\alpha} &=& - \ln \left( \frac{b}{b_{cr}} - 1 \right) + \ln[216(7-4\sqrt{3})] - \pi \, = \nonumber\\
&=& - \ln \frac{b-3\sqrt{3}M}{648\sqrt{3}(7-4\sqrt{3})M} - \pi   = - \ln \frac{(b-3\sqrt{3}M)(7+4\sqrt{3})}{648\sqrt{3}M} - \pi \, .
\end{eqnarray}

In the strong deflection approximation, the deflection angle diverges
logarithmically while the impact parameter approaches its critical value \cite{Bozza2001}. It was shown in \cite{Bozza2002} that for any spherically symmetric space-time the deflection angle diverges logarithmically when the impact parameter approaches its critical value.

In the case of an unbound motion of massive particle near the black hole, the deflection angle is also written via integral and can be expressed via elliptic integrals (see, for example, \cite{Tsupko2014}). In case of weak deflection ($\hat{\alpha} \ll 1$) the deflection angle is described by the formula (see, for example, \cite{MTW}):
\begin{equation} \label{alpha-small}
\hat{\alpha} = \frac{R_S}{b} \left( 1 + \frac{1}{v^2} \right).
\end{equation}
Here $b$ and $v$ are the impact parameter and velocity of incident test particles, respectively. Analytical formulae for the case of a massive particle motion near Schwarzschild black hole, similar to (\ref{vacuum-alpha-R}), (\ref{vacuum-alpha-b}), have been obtained by Tsupko \cite{Tsupko2014}, in the limit of a strong deflection angle ($\hat{\alpha} \gg 1$), when the impact parameter approaches its critical value.


Photons from a distant source which undergo one or several loops around the central object (lens), and then go to observer form images, which are called relativistic images \cite{Virbhadra2000} (see Fig.\ref{figure-prd2}). When the lens, source and the observer are situated on one line, these images form concentring rings around the lens \cite{BKTs2008}, see Fig.\ref{figure-rings}. In the opposite case the relativistic images are represented by two infinite sequences of 'spots' situated on one line at both sides of the lens, and the closest approach distance of photons is converging to the radius $3M$. Using an exact expression for the deflection angle, Virbhadra and Ellis \cite{Virbhadra2000} calculated numerically the positions and the magnifications of the relativistic images for the Schwarzschild space-time. Frittelli, Kling, Newman \cite{Frittelli2000} considered the exact lens equation for the Schwarzschild metric, and obtained solutions in the form of an integral expressions. Exact gravitational lens equation in spherically symmetric and static space-times was also considered in Perlick \cite{Perlick2004a}. A review and comparison of different approximate lens equations is given in \cite{Bozza2008}.

Relativistic images can be studied analytically with using of the deflection angle in the strong deflection limit. To make several revolutions around black hole, the distance of the closest approach must be very close to $r_M=3M$, and formulae (\ref{vacuum-alpha-R}) and (\ref{vacuum-alpha-b}) provide very good accuracy. Bozza et al \cite{Bozza2001} applied analytical expression for the investigation of the relativistic images, finding the positions and the magnifications of the relativistic images in the Schwarzschild metric. Relativistic rings (see Fig.~\ref{figure-rings}) for a Schwarzschild black hole lens were considered in more details in \cite{BKTs2008}.

Recent decade  lensing beyond the weak deflection limit was studied in many papers, where the approaches described above were applied to  different problems \cite{Virbhadra2000}, \cite{Bozza2001}, \cite{Perlick2004a}, \cite{TsBK2013}, \cite{BKTs2008}, \cite{Perlick2010},  \cite{Bozza2002}, \cite{Perlick2004review}, \cite{Bozza2008} - \cite{Cite3}. Relativistic images in vacuum for different types of metric \cite{Perlick2004review}, \cite{Perlick2010}, or using alternative theories of gravitation \cite{Bozza2010} have been studied.


\section{General formulae for the photon deflection angle in the Schwarzschild metric in presence of plasma}

In this section we derive the exact expression for the deflection angle in the case of a photon motion in the Schwarzschild
metric with a spherically symmetric distribution of plasma. We use spherical coordinates $(r, \theta, \varphi)$ and the gravitational
field is not supposed to be weak. Indices are
\begin{equation}
i,k = 0,1,2,3; \quad \alpha, \beta = 1,2,3 \; (r, \theta, \varphi) ; \quad \mbox{signature} \; \; \{-,+,+,+\} .
\end{equation}

Let us consider a static space-time with the Schwarzschild metric ($R_S=2M$, $G=c=1$):
\begin{equation}
ds^2 = g_{ik} dx^i dx^k = - A(r) \, dt^2 + \frac{dr^2}{A(r)} + r^2
\left( d \theta^2 + \sin^2 \theta d \varphi^2 \right),  \;  A(r) = 1 - \frac{2M}{r} .
\label{metric}
\end{equation}
Let us consider, in this gravitational field, a static inhomogeneous plasma with a refraction index
\begin{equation} \label{plasma-n}
n^2 = 1 - \frac{\omega_e^2}{[\omega(r)]^2} \, , \quad \omega_e^2 = \frac{4 \pi e^2 N(r)}{m} \, .
\end{equation}
Here $N(r)$ is the electron concentration in plasma.

The general relativistic geometrical optics in a curved space-time, in a dispersive medium with an angular isotropy of the refraction index, was developed by Synge
\cite{Synge}, see also \cite{Bicak}. Let us assume that the space-time $g_{ik}$ is given, and the refraction index $n$ of an isotropic transparent medium is given by a scalar function of $x^\alpha$, and the photon frequency $\omega(x^\alpha)$. The Synge's method is based on the so called medium-equation:
\begin{equation} \label{eq-medium-general}
n^2 = 1 + \frac{p_i p^i}{\left(p_k V^k \right)^2} \, .
\end{equation}
This formula connects the components of the photon momentum 4-vector (the photon energy and 3-vector of the photon momentum) and refraction index $n$. In this equation the metric $g_{ik}$ and the refraction index $n$ are assumed to be known, a phase speed $w=1/n$ and the frequency $\omega(x^\alpha)$ are measured in the local rest frame, $V^k$ is 4-velocity of medium. There is also a relation for the photon frequency \cite{Synge}
\begin{equation}
p_i V^i  = - \hbar \omega(x^\alpha) \, ,
\end{equation}
where $\hbar$ is the Planck constant.

For a static medium in a static gravitational field, we have the medium-equation in the form \cite{Synge}:
\begin{equation} \label{eq-medium}
n^2 = 1 + \frac{p_i p^i}{\left(p_0 \sqrt{-g^{00}}\right)^2} \, ,
\end{equation}
and
\begin{equation} \label{Synge}
p_0 \sqrt{-g^{00}} = - p^0 \sqrt{-g_{00}} = - \hbar \omega(x^\alpha) \, ,
\end{equation}
 At infinity, in a flat space-time, we have:
\begin{equation} \label{Synge2}
p_0 = - p^0  = - \hbar \omega \, ,
\end{equation}
where $\omega \equiv \omega(\infty)$.

In order to apply Hamiltonian method, we rewrite the medium-equation (\ref{eq-medium}) in the form
\begin{equation}
H(x^i,p_i) = 0 \, , \label{restriction}
\end{equation}
where
\begin{equation}
H(x^i,p_i) = \frac{1}{2} \left[ g^{ik} p_i p_k - (n^2-1) \left(p_0 \sqrt{-g^{00}}\right)^2 \right]  \, .
\end{equation} \label{H-alpha}
Here we define the scalar function $H(x^i,p_i)$ depending on $x^i$ and $p_i$.

The trajectories of photons, in presence of a gravitational
field, may be obtained from the variational principle \cite{Synge}
\begin{equation} \label{var-princ}
\delta \left(\int p_i \, dx^i\right) = 0 \, ,
\end{equation}
with the restriction (\ref{restriction}). The variational principle (\ref{var-princ}), with the
restriction $H(x^i,p_i)=0$, leads to the following system of Hamiltonian differential equations \cite{Synge}:
\begin{equation}
\label{D-Eq} \frac{dx^i}{d \lambda} = \frac{\partial H}{\partial p_i}  \, , \; \; \frac{dp_i}{d \lambda} = - \frac{\partial
H}{\partial x^i} \, ,
\end{equation}
with the parameter $\lambda$ changing along the light trajectory.

In the case of a plasma with the refractive index
(\ref{plasma-n}), the condition (\ref{restriction}) can be reduced,
with using of (\ref{plasma-n}) and (\ref{Synge}), to the form
\begin{equation}
H(x^i,p_i) = \frac{1}{2} \left[ g^{ik} p_i p_k + \omega_e^2 \hbar^2 \right] = 0 \, . \label{H-definition}
\end{equation}
From (\ref{D-Eq}) we obtain the system of equations for the space
components $x^\alpha$, $p_\alpha$:
\begin{equation} \label{eq-motion-x-general}
\frac{dx^\alpha}{d \lambda} = g^{\alpha \beta} p_\beta \, ,
\end{equation}
\begin{equation} \label{eq-motion-p-general}
\frac{dp_\alpha}{d \lambda} = -\frac{1}{2} \, g^{ik}_{,\alpha} p_i p_k - \frac{1}{2} \, \hbar^2 \left( \omega_e^2 \right)_{,
\alpha} \, .
\end{equation}
Equation for the time component $dp_0/d \lambda = 0$ in the static metric (\ref{metric}) leads to $p_0 = $ const\, along the trajectory. The equation
(\ref{Synge}) at infinity has the form (\ref{Synge2}), so we find that the constant $p_0$ equals to $p_0= - \hbar \omega$.

Let us find the equation of the trajectory, and the photon deflection angle, for a motion in the equatorial plane $\theta = \pi /2$ of the metric (\ref{metric}). Then $\sin^2 \theta=1$, and components of the metric have a form
\begin{equation}
g_{rr} = \frac{1}{A(r)}, \; g^{rr} = A(r), \; g_{00} = -A(r), \; g^{00} = - \frac{1}{A(r)},
\label{met1}
\end{equation}
\begin{equation}
g_{\varphi \varphi} = r^2, \quad g^{\varphi \varphi} =
\frac{1}{r^2}, \quad g_{\theta \theta} = r^2, \quad g^{\theta
\theta} = \frac{1}{r^2}.
\label{met2}
\end{equation}
From the equation for $\theta$ we have  from (\ref{eq-motion-x-general})

\begin{equation}
\frac{d \theta}{d \lambda} = g^{\theta \theta} p_\theta \, , \;\; \mbox{ or } \;\; \frac{d \theta}{d \lambda} = p^\theta \,.
\end{equation}
It follows that for a motion in the plane $\theta=\pi/2$ we have $p_\theta = p^\theta = 0$. From the equation  (\ref{eq-motion-p-general}) for $p_\varphi$ it
follows  $p_\varphi = \mbox{const}$. Without a loss of generality we can assume that $p_\varphi>0$.
The equations for $r$ and $\varphi$ from (\ref{eq-motion-x-general}) may be written as
\begin{equation}
\frac{dr}{d\lambda} = g^{rr} p_r \, , \quad  \frac{d\varphi}{d\lambda} = g^{\varphi \varphi} p_\varphi \, .
\end{equation}
Substituting the components of metric we obtain:
\begin{equation} \label{dphidr-prelim}
\frac{d \varphi}{dr} = \frac{p_\varphi}{r^2 } \, \frac{1}{p_r A(r)} \, .
\end{equation}
The second multiplier in the right hand side of this equation can be expressed from the medium equation for plasma
(\ref{H-definition}), which may be written in the form
\begin{equation}
g^{rr} p_r^2  + g^{\varphi \varphi} p_\varphi^2 + g^{00} p_0^2 + \hbar^2 \omega_e^2(r) = A(r) p_r^2  + \frac{p_\varphi^2}{r^2}  - \frac{p_0^2}{A(r)}  + \hbar^2 \omega_e^2(r) = 0 \, .
\end{equation}

We have then
\begin{equation} \label{prAr}
p_r A(r) = \pm \sqrt{p_0^2 - A(r)  \left( \frac{p_\varphi^2}{r^2} + \hbar^2 \omega_e^2(r) \right)} \, .
\end{equation}
Substituting (\ref{prAr}) into (\ref{dphidr-prelim}), we obtain the equation for trajectory of photon:
\begin{equation} \label{eq-traekt-general}
\frac{d \varphi}{dr} = \pm \frac{p_\varphi}{r^2 } \, \frac{1}{\sqrt{p_0^2 - A(r)  \left( \frac{p_\varphi^2}{r^2} + \hbar^2
\omega_e^2(r) \right)}} \, .
\end{equation}
Assume that a photon moves in such a way that its $\varphi$-coordinate increases. Then 'plus' sign in (\ref{eq-traekt-general}) corresponds to the motion with the coordinate $r$ also increasing, and the 'minus' sign to the motion with $r$ decreasing.

For a photon which moves from the infinity to the distance of the closest approach $R$ (minimal value of the coordinate $r$), and then to infinity, we have a change of angular coordinate as
\begin{equation}
\Delta \varphi = - \int \limits_\infty^R \frac{p_\varphi}{r^2 } \, \frac{dr}{\sqrt{p_0^2 - A(r)  \left(\frac{p_\varphi^2}{r^2} + \hbar^2 \omega_e^2(r) \right)} }  + \int \limits_R^\infty \frac{p_\varphi}{r^2 } \, \frac{dr}{\sqrt{p_0^2 - A(r)  \left(\frac{p_\varphi^2}{r^2} + \hbar^2 \omega_e^2(r) \right)} }   =
\end{equation}
$$
= 2 \int \limits_R^\infty \frac{p_\varphi}{r^2 } \, \frac{dr}{\sqrt{p_0^2 - A(r)  \left(\frac{p_\varphi^2}{r^2} + \hbar^2 \omega_e^2(r) \right)} } \, .
$$
The motion along a straight line corresponds to the change of the angular coordinate $\Delta \varphi = \pi$.
Then the deflection angle may be written as
\begin{equation} \label{angle-general}
\hat{\alpha} = 2 \int \limits_R^\infty \frac{p_\varphi}{r^2 } \, \frac{dr}{\sqrt{p_0^2 - A(r)  \left(
\frac{p_\varphi^2}{r^2} + \hbar^2 \omega_e^2(r) \right)} }  - \pi \, .
\end{equation}

The deflection angle (\ref{angle-general}) depends on a mass of the central body $M$, a
distribution of plasma $N(r)$, represented by $\omega_e(r)$, and on the parameters $R$, $p_0$ and $p_\varphi$. These parameters are connected by the boundary condition. The point $r=R$ is a turning point,
therefore in this point: $dr/d \lambda = 0$ and $p_r = 0$. Using (\ref{prAr}), we have in this point
\begin{equation} \label{boundary-general}
p_0^2 = A(R)  \left( \frac{p_\varphi^2}{R^2} + \hbar^2 \omega_e^2(R) \right) \, .
\end{equation}
Therefore only two parameters from \{$R$, $p_0$, $p_\varphi$\} are independent, while the third one is expressed through two others. The parameter $p_0$ represents the photon energy at infinity. It is convenient to exclude $p_\varphi$, and to derive the angle $\hat{\alpha}$ as a function of $p_0 = - \hbar \omega$ and $R$.  We have from (\ref{boundary-general})
\begin{equation}
p_\varphi^2 = R^2 p_0^2 \left( \frac{1}{A(R)} - \frac{\omega_e^2(R)}{\omega^2}  \right) \, .
\end{equation}
Substituting $p_\varphi^2$ in (\ref{angle-general}) and using notation of Perlick \cite{Perlick2000}
\begin{equation} \label{perlick-h}
h(r) = r \sqrt{ \frac{1}{A(r)} - \frac{\omega_e^2(r)}{\omega^2} } = r \sqrt{ \frac{r}{r-2M} - \frac{\omega_e^2(r)}{\omega^2} } \, ,
\end{equation}
we obtain the equation of the trajectory in the Schwarzschild space-time
\begin{equation} \label{perlick-trajectory}
\frac{d \varphi}{dr} = \pm \frac{1}{\sqrt{r(r-2M)}\sqrt{\frac{h^2(r)}{h^2(R)} - 1 }} \, ,
\end{equation}
and the deflection angle of the photon moving from infinity to the central object and then to infinity
\begin{equation} \label{perlick-angle}
\hat{\alpha} = 2 \int \limits_R^\infty \frac{dr}{\sqrt{r(r-2M)}\sqrt{\frac{h^2(r)}{h^2(R)} - 1 }} - \pi \, .
\end{equation}
This expression for $\hat\alpha$ was derived earlier in \cite{Perlick2000} by an another way. At given $M$ and $\omega_e(r)$, the deflection angle is determined by the closest approach distance $R$ and the photon
frequency at infinity $\omega$. This formula allows us to calculate the deflection angle of the photon moving in the Schwarzschild metric in presence of a spherically symmetric distribution of plasma.

\section{Gravitational deflection of light in homogeneous plasma: exact expression. Analogy between a photon in a homogeneous plasma and a massive particle in a vacuum.}

Let us consider a homogeneous plasma, $\omega_e(r)=\omega_e =$ const.
Rewriting the equation of the trajectory (\ref{eq-traekt-general}) and the expression for deflection angle (\ref{angle-general}), introducing
notations of $E$ and $L$
\begin{equation} \label{definition-E-L}
\frac{-p_0}{\hbar \omega_e} = \frac{\omega}{\omega_e} = E > 1 \, , \quad \frac{p_\varphi}{\hbar \omega_e} = L > 0 \, ,
\end{equation}
we obtain the equation of the trajectory as
\begin{equation} \label{trajectory-EL}
\frac{d \varphi}{dr} = \pm \frac{L}{r^2} \frac{1}{\sqrt{E^2 - A(r) \left( 1 + \frac{L^2}{r^2} \right) }} \, ,
\end{equation}
and the formula for deflection angle as
\begin{equation} \label{angle-EL}
\hat{\alpha} =  2 \int \limits_R^\infty \frac{L}{r^2} \frac{dr}{\sqrt{E^2 - A(r) \left( 1 + \frac{L^2}{r^2}  \right) }} - \pi \, .
\end{equation}
The closest approach distance $R$ and the parameters $E$ and $L$ are connected at the point $r=R$ by the following boundary condition
\begin{equation} \label{boundary-cond-EL}
E^2 = A(R) \left( 1 + \frac{L^2}{R^2} \right) \, .
\end{equation}
The trajectory and the deflection angle are thus completely determined by any two parameters from \{$R$,  $E$, $L$\}, with the
third parameter being expressed through (\ref{boundary-cond-EL}). Remind, that in vacuum the photon motion
is determined by only one parameter, either $R$, or $b$, which are uniquely connected with each other, see (\ref{b-and-R}).
Transition to the vacuum from (\ref{boundary-cond-EL}) can be done by tending $E \rightarrow \infty$ and $L \rightarrow \infty$ (what corresponds to $\omega_e=0$) and introducing the impact parameter as $b^2 = L^2/(E^2-1)$, as in the analogous case of the massive particle \cite{MTW}.

Plasma has unique dispersive properties, that leads to interesting analogy between motion of photon in homogeneous plasma and massive particle in vacuum. In the paper of Kulsrud and Loeb \cite{Kulsrud-Loeb} (see also papers of Broderick and Blandford \cite{Brod-Blandford1}, \cite{Brod-Blandford2}) it was shown that in the homogeneous plasma the photon wave packet moves like a particle with a velocity equal to the group velocity of the wave packet
\begin{equation}
v_{pl} = v_{group} =\sqrt{1-\frac{\omega_e^2}{\omega^2}} \, ,
\end{equation}
with a mass equal to the plasma frequency (multiplied by $\hbar$)
\begin{equation}
m_{pl} = \hbar \omega_e \, ,
\end{equation}
and with an energy equal to the photon energy
\begin{equation}
E_{pl} = \hbar \omega \, .
\end{equation}

If we define $E$ as an energy at infinity per unit rest mass, and $L$ as the angular momentum per unit rest mass, then equations (\ref{trajectory-EL}) and (\ref{angle-EL}) describe the trajectory and the deflection angle of a massive particle with the rest mass $\hbar \omega_e$ in the vacuum, see \cite{Kulsrud-Loeb, MTW, Weinberg, Zeld-Novikov}).

As for massive particle in vacuum, bound elliptic orbits of the photon in homogeneous plasma are also possible. Discussion of gravitational binding of the photon in homogeneous plasma is discussed in  \cite{Kulsrud-Loeb},\cite{TsBK2013}.

The deflection angle (\ref{angle-EL}) can be expressed via elliptic integrals. Due to described analogy, the expression for the deflection angle of a massive particle in vacuum is identical to (\ref{angle-EL}), with an appropriate changes in the physical sense of $E$ and $L$. The motion of a massive
particles has been extensively studied, see \cite{Hagihara1931,Darwin1959,Darwin1961,Bogorodsky1962, M-Pleb1962,Metzner1963,
Zeld-Novikov,MTW,Chandra,MiroRodriguez1,MiroRodriguez2}. In particular, the deflection angle of a
massive particle can be expressed via elliptic integrals. Usually the deflection angle is rewritten as an integral of
the polynomial of the third order in a radicand, see \cite{TsBK2013} for details. The form of the expression via elliptic integrals is determined by the roots of this polynomial. In a general case, these three roots are different,
and supposed to be found numerically (or by cumbersome solution of cubic equation, see, for example, \cite{Bronstein}), for the given external parameters $E$ and $L$. Depending on $E$ and $L$, the roots have
different relations between each other, and the expression of the deflection angle via elliptic integrals have therefore different
form. In the newtonian limit these forms are expressed analytically, and are  related to hyperbolic, parabolic, and elliptic motion of a massive particle around a gravitating center (Kepler problem), see \cite{LLmech}. The principal feature is that we cannot write the deflection angle as an explicit function of $E$ and $L$. For given $E$ and $L$, one has to obtain numerically the roots of the polynomial before being able to
compute the deflection angle (see, for example, \cite{M-Pleb1962}, \cite{Metzner1963}).

In the paper \cite{TsBK2013} we suggested to use, as independent, another pair of parameters: $R$ and $E$. Then one root of the polynomial is simply $1/R$, and another two roots can be expressed analytically via $R$ and $E$. This approach is convenient because we can perform all
calculations analytically and obtain the deflection angle in the form of elliptic integrals as an explicit function of two
parameters $R$ and $E$.  A similar approach is usually applied for the deflection of photons in vacuum, where an analogous polynomial is expressed via $R$.

The problem is that $R$ is not an {\it ad hoc} value and thus should not be chosen as an external parameter in applications. For example, for gravitational lens theory we need dependence of the deflection angle on the impact parameter.
This difficulty can be avoided in strong deflection limit, where it becomes possible to express  the deflection angle in terms of
$b$ and $E$, which is convenient for applications (see Section 8 and \cite{TsBK2013} for details).

The deflection angle $\hat{\alpha}$ for the photon in the homogeneous plasma can be written in the form \cite{TsBK2013}
\begin{equation} \label{alpha-exact1}
\hat{\alpha} = 4 \sqrt{\frac{R}{Q}} \, F(y, k)  -  \pi  \, , \quad \mbox{with} \quad y = \sqrt{  \frac{8MQ}{( 6M-R+Q )( R-2M+Q )} }  \, ,
\end{equation}
\[
k = \sqrt{ \frac{6M-R+Q}{2Q} } \, , \quad  Q^2 = (R-2M)^2 + 8M (R-2M) \, \frac{1}{1+ \frac{2M}{R(E^2-1)}} \, .
\]
Here $F(z,k)$ is an elliptic integral of the first kind \cite{Korn, Gr-Ryzhik}:
\begin{equation}
F(z,k) = \int \limits_0^z \frac{dx}{\sqrt{(1-x^2)(1-k^2 x^2)} } =  \int \limits_0^\varphi \frac{d \theta}{\sqrt{1 - k^2 \sin^2 \theta}} =  \tilde{F} (\varphi, k) \, ,
\end{equation}
$$
F(\sin \varphi, k) \equiv \tilde{F} (\varphi, k) \, ,  \;  x = \sin \theta \, , \;  z=\sin \varphi \, .
$$

The expression $\hat{\alpha}$ can be also written in another form \cite{TsBK2013}
\begin{equation} \label{alpha-exact2}
\hat{\alpha} = 4 \sqrt{\frac{R}{Q}} \, \left[ F(1,k) - F(z,k) \right]   -  \pi.
\end{equation}
Here $F(1,k) = \tilde{F}(\pi/2,k) = K(k)$ is the complete elliptic integral of the first kind, and
\begin{equation}
\label{z2}
z^2 = \frac{2M+Q-R}{6M+Q-R} \, .
\end{equation}

This formula is written in the same form as the expression for the vacuum
deflection angle, see \cite{MTW}, \cite{Chandra}, \cite{IyerPetters}, \cite{BKTs2008}. The difference between plasma and vacuum
formulae is only in the expression for $Q$. When $E$ goes to infinity, what corresponds to high energy photons, for which plasma effects are negligible, the expression for $Q$ in (\ref{alpha-exact1}) goes to  $Q^2 =
(R-2M)(R+6M)$, and the formula (\ref{alpha-exact2})  transforms into the formula for the vacuum deflection.
The expression (\ref{alpha-exact1}) for the vacuum case with $Q^2 = (R-2M)(R+6M)$ is written as
\cite{Darwin1959}

\begin{equation}
\hat{\alpha} = 4 \sqrt{\frac{R}{Q}} \, F(y, k)  -  \pi   \,  ,
\end{equation}
\begin{equation}
\mbox{with} \quad y = \sqrt{  \frac{2Q}{3R-6M+Q} }  \, , \quad k = \sqrt{ \frac{6M-R+Q}{2Q} }  \, .
\end{equation}

The proposed novel technique can be also applied for a massive particle in vacuum, due to its analogy to
a photon in plasma. In paper \cite{Tsupko2014} we suggest to use either the pair ($R$, $L$) or ($R$, $E$) as an independent pair of parameters, instead of ($E$, $L$). In that paper we obtain the deflection angle in the form of elliptic integrals as an explicit function of the two parameters ($R$, $L$) or ($R$, $E$). We present for the first time the analytical formulae for the deflection angles of massive particles in the strong deflection limit ($\hat{\alpha} \gg 1$) as an explicit function of external parameters at infinity, ($b$, $L$) or ($b$, $E$). Parameters at infinity are simply related with each other, so it is possible to calculate the deflection angle with our formulae in very different situations. For example, our formulae allows one to calculate analytically the deflection angle of particle with given impact parameter and velocity at infinity as initial conditions, beyond usual small deflection case.

\section{Gravitational lensing in the limit of a small deflection in presence of a homogeneous plasma: analytic solution}
\label{section-uniform}

Exact expression for deflection angle in presence of Schwarzschild gravity and plasma (\ref{perlick-angle}) is manifestation of simultaneous and mutual action of different physical effects: gravity, refraction, dispersion. This angle (\ref{perlick-angle})  can not be calculated analytically. For analysis and applications of this result it is convenient to consider some particular cases with approximations. Considering case of small total deflection $\hat{\alpha} \ll 1$, we have advantages to see clearly which physical reasons give contributions to different parts of the total deflection. In particular, we will be able to discuss the difference of this angle and known, commonly used, formulae; see this and next sections.

In the case of a homogeneous plasma the refractive index does not depend on space coordinates explicitly, so the refractive terms in deflection are absent. Therefore, in homogeneous case, we can consider this angle $\hat{\alpha}$ as the gravitational deflection $\alpha_{grav}$ in given medium (plasma), and write
\begin{equation}
\hat{\alpha} = \alpha_{grav}.
\end{equation}
We use notation $\alpha_{grav}$ here to emphasize that in homogeneous case the total deflection $\hat{\alpha}$ consists of only the gravitational deflection (compare with the next section).

Let us consider a situation when the closest approach distance is much larger than the Schwarzschild radius, $R \gg M$ ($R_S=2M$). It means also that $r \gg M$ (during the motion the $r$-coordinate changes from $R$ to infinity), $b \gg M$, and the resulting deflection angle is small, $\hat{\alpha} \ll 1$. Expanding the exact formula (\ref{perlick-angle}) with $\omega_e^2 =$ const, we obtain the deflection angle in the form (see Appendix 1 for details):
\begin{equation} \label{angle-plasma-R}
\hat{\alpha} = \alpha_{grav} = \frac{2M}{R} \left( 1 + \frac{1}{1 - \omega_e^2/\omega^2} \right) ,
\end{equation}
\[
\mbox{or, in usual units,} \; \; \hat{\alpha} = \frac{2GM}{c^2 R} \left( 1 + \frac{1}{1 - \omega_e^2/\omega^2} \right), \;\; R_S = \frac{2GM}{c^2}.
\]

For gravitational lensing the dependence of angles on impact parameter $b$ is needed. In approximation $R \gg M$ the difference between $R$ and $b$ is negligible, so we can just substitute $R \simeq b$ into (\ref{angle-plasma-R}), obtaining
\begin{equation} \label{angle-plasma-b}
\hat{\alpha} = \frac{2M}{b} \left( 1 + \frac{1}{1 - \omega_e^2/\omega^2} \right) .
\end{equation}

The deflection angle (\ref{angle-plasma-b}) was directly derived in a simpler way in \cite{BKTs2009}, \cite{BKTs2010}. To do this, the Synge's equations have been written in Cartesian coordinates. We have considered the photon with unperturbed trajectory in a homogeneous plasma as a straight line parallel to $z$-axis with an impact parameter $b$, and have found the deflection in the approximation of small perturbations, by the integration along the straight trajectory. Compare: the vacuum gravitational deflection in the form $2R_S/R$ was derived by expansion the integral of the exact deflection angle \cite{LL2}, and the deflection in the form $2R_S/b$ was derived also by the integration of small deflections along the trajectory \cite{MTW}.

Formula (\ref{angle-plasma-b}) is valid only for $\omega > \omega_e$, because the waves with $\omega< \omega_e$ do not propagate in the plasma. If $\omega \simeq \omega_e$, then gravitational deflection in plasma can be much larger than vacuum gravitational deflection, $\hat{\alpha} \gg 2R_S/b$ (condition $\hat{\alpha} \ll 1$ must be satisfied too).

Formula (\ref{angle-plasma-b}) does not imply that $\omega_e/\omega \ll 1$. If it is supposed, we can rewrite the deflection (\ref{angle-plasma-b}) as:
\begin{equation}
\hat{\alpha} = \alpha_{grav} = \alpha_{einst} + \alpha_{add} = \frac{2R_S}{b} + \frac{\omega_e^2}{2\omega^2} \frac{2R_S}{b} .
\end{equation}
Here we denote the vacuum gravitational deflection as $\alpha_{einst}$ and additional correction to the gravitational deflection connected with plasma presence as $\alpha_{add}$. We emphasize that the correction $\alpha_{add}$ is presented also in a non-homogeneous plasma, but has a more complicated form.

The presence of plasma increases the gravitational deflection angle. In homogeneous plasma photons of smaller frequency, or larger wavelength, are deflected by a larger angle by the gravitating center. The effect of difference in the gravitational deflection angles is significant for longer wavelengths, when $\omega$ is approaching $\omega_e$. That is possible only for the radio waves. Therefore, the gravitational lens in plasma acts as a radiospectrometer \cite{BKTs2009}.

We should use formula (\ref{main-res}) when we consider gravitational lensing of radiowaves by point or spherical body in presence of homogeneous plasma. This effect has a general relativistic nature, in combination with the dispersive properties of plasma. We should also emphasize that the plasma is considered here like the medium with a given index of refraction, and this formula does not take into account gravitation of particles of plasma.

\begin{figure}
\centerline{\hbox{\includegraphics[width=0.4\textwidth]{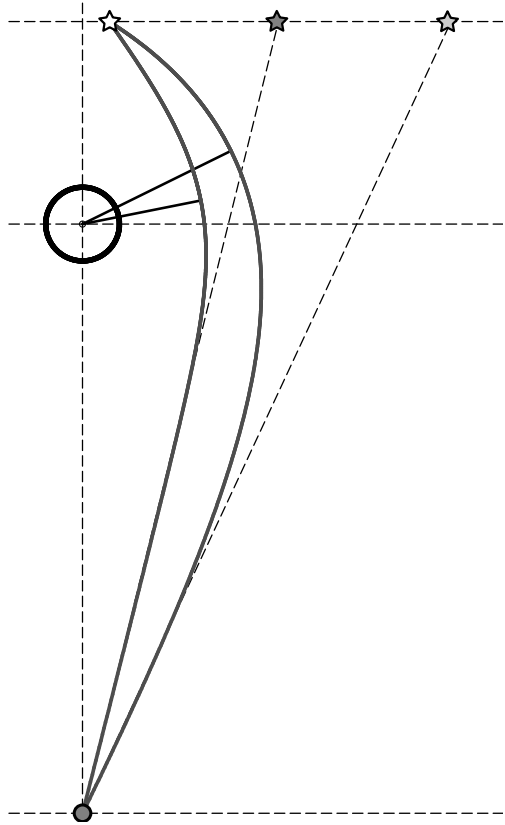}}} \caption{
Gravitational lensing in plasma, scheme of formation of the primary images of different frequencies. Two light rays which have different frequencies are shown. To reach the observer these rays are deflected by different angles. Therefore angular positions of images are different at different wavelengths.} \label{figure-prd3}
\end{figure}

The observational effect of the frequency dependence may be represented on the example of the Schwarzschild point-mass lens. Instead of two concentrated images with complicated spectra, we will have two expanded line images, formed by the photons with different frequencies, which are deflected by different angles (Fig.\ref{figure-prd3}). Estimations of an angular difference of positions of images in different bands can be found in our paper \cite{BKTs2010}. To see the difference in angular position of images, we should compare radio and optical observations of images. For optical frequencies the effect of plasma presence is negligible, and positions of images in this case can be calculated with the vacuum formulae. The difference between angular separations of images in vacuum and in plasma $\Delta\theta_0$, produced by the same lens configuration with point-mass lens, is equal
to
\begin{equation} \label{difference}
\frac{\Delta \theta_0}{\theta_0} = \frac{1}{4} \frac{\omega_e^2}{\omega^2} \simeq
2.0 \cdot 10^7 \, \frac{N_e}{\nu^2} ,
\end{equation}
where $\nu$ is the photon frequency in Hz, $\omega = 2 \pi \nu$, $\theta_0$ is angular radius of Einstein ring \cite{BKTs2010}. This formula is written for homogeneous plasma, for non-homogeneous plasma see analytical formulae in next section of present paper and modelling and estimations in \cite{Mao2014}. We can also compare two radio observations with the different frequencies. For one of two frequencies the effect of angular shifting is bigger than for another frequency, so we may observe the difference in positions of images of different frequencies.

Gravitational lensing also leads to magnification of source. This means that the flux of image is bigger or smaller than the flux of source, different images have different magnifications. In observations we don't know the intrinsic flux of source and its spectrum, but we can compare the fluxes of images in different frequencies. The ratios of the fluxes of images in different bands should be the same, if we consider lensing in vacuum, because gravitational deflection in vacuum is achromatic.

The magnification is determined by the deflection law \cite{GL}. In the case of radio lensing in the plasma we have another formula for deflection instead of the Einstein angle, so formulae for magnification will be different. It leads to difference of magnigications of different images in different bands, when the light propagates in regions with different plasma density. By-turn it leads to difference of ratios of the fluxes of images in different bands (see details in \cite{BKTs2010}). It is another prediction of the model of gravitational radiospectrometer for observation. We should mention that Thompson scattering and absorption during propagation of radiation through the plasma can significantly change flux and complicate phenomena of lensing magnification.

\begin{figure}
\centerline{\hbox{\includegraphics[width=0.9\textwidth]{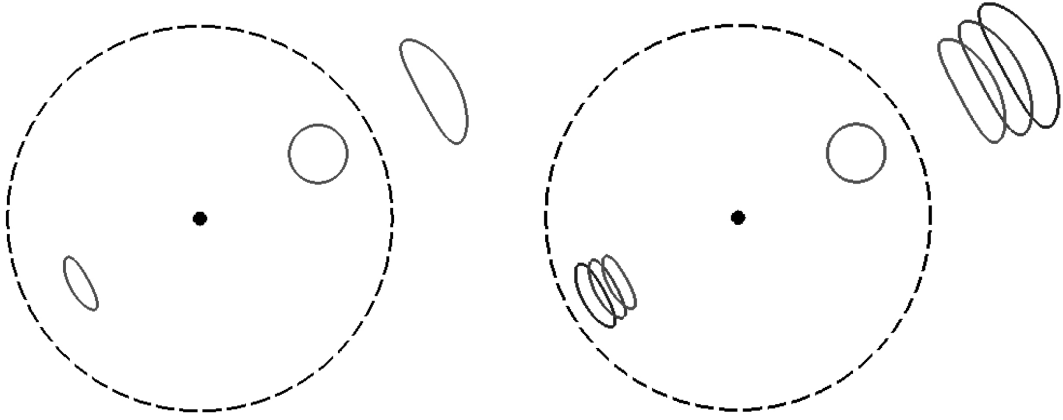}}} \caption{
Comparison of lensing by point source in vacuum (left) and in plasma (right). The photon
trajectories and deflection angles in vacuum don't depend on the photon frequency (or energy). Therefore lensing in vacuum is achromatic, and the Schwarzschild point-mass lens leads to two images with complicated spectra, formed by the photons of different frequencies which undergo deflection by the same angle. If lens is surrounded by plasma, the light rays of different frequencies have different deflection angles. The images become diffused and divide into many images formed by photons of different frequencies. Angular separation between images of different frequencies take place both in homogeneous and non-homogeneous plasma and is subject for observation.} \label{figure-spectr}
\end{figure}

\section{Gravitational lensing in the limit of small deflection, in presence of a non-homogeneous plasma}
\label{section-nonuniform}

Discussion above is valid for the Schwarzschild lens in a homogeneous plasma. In the case of plasma non-homogeneity there is also the refractive deflection, $\alpha_{refr}$. Our approach for gravitational lensing in plasma developed in \cite{BKTs2009}, \cite{BKTs2010} allows us consider two effects simultaneously: the gravitational deflection in plasma which is different from the Einstein angle, and the refraction connected with the plasma inhomogeneity which does not depend on the gravity. In the paper \cite{BKTs2010} we have derived formulae for the deflection angle by the spherical body or spherically distributed gravitating matter in presence of a homogeneous and non-homogeneous plasma, with taking into account the gravitation deflection together with the refractive deflection.

In applications the usual vacuum gravitational deflection $\alpha_{einst}$ gives the major contribution to the total deflection $\hat{\alpha}$, and all effects connected with plasma presence are significally smaller: $\alpha_{add} \ll \alpha_{einst}$ and $\alpha_{refr} \ll \alpha_{einst}$.

Additional correction $\alpha_{add}$ to the gravitational deflection caused by plasma (see previous section) is usually smaller than correction due to refraction $\alpha_{refr}$, but there are situations when these corrections can be of the same order, see \cite{BKTs2010}. Let us consider situation when correction to gravitational deflection is smaller than refractive deflection ($\alpha_{add} \ll \alpha_{refr}$), and obtain the total deflection angle (in weak deflection approximation) as a sum of vacuum gravitational deflection $\alpha_{einst}$ and refractive deflection $\alpha_{refr}$. In this approximation we neglect $\alpha_{add}$ and consider that the gravitational deflection is equal to vacuum gravitational deflection ($\alpha_{grav} \simeq \alpha_{einst}$).

Expanding formula for exact deflection (\ref{perlick-angle}), we obtain approximate formula in the case of a weak deflection (see Appendix 2 for details):
\begin{equation} \label{angle-nonhomogen}
\hat{\alpha} =  \alpha_{einst} + \alpha_{refr} \, ,
\end{equation}
\begin{equation}
\alpha_{einst} = \frac{4M}{R} = \frac{2R_S}{R} \, ,
\end{equation}
\begin{equation} \label{angle-refr1}
\alpha_{refr} =  \frac{R K_e}{\omega^2} \int \limits_R^\infty \frac{1}{\sqrt{r^2-R^2}} \frac{dN(r)}{dr} \, dr \, .
\end{equation}

For gravitational lensing the dependence of angles on the impact parameter $b$ is needed. In papers \cite{BKTs2009} and \cite{BKTs2010} we have derive $\hat{\alpha}$ using Cartesian coordinates. We have considered the photon with the unperturbed trajectory as a straight line parallel to $z$-axis, with the impact parameter $b$. We have obtained the gravitational deflection at given $b$ as
\begin{equation} \label{einst-b}
\alpha_{einst} = \frac{4M}{b} = \frac{2R_S}{b} \, ,
\end{equation}
and refractive deflection as \cite{BKTs2009}
\begin{equation} \label{angle-refr2}
\alpha_{refr} =  \frac{K_e}{\omega^2}   \int \limits_0^\infty \frac{\partial N}{\partial b} \, dz
\end{equation}
or \cite{BKTs2010}
\begin{equation} \label{angle-refr3}
\alpha_{refr} =  \frac{K_e b}{\omega^2}   \int \limits_0^\infty \frac{1}{r} \frac{dN}{dr} \, dz \, .
\end{equation}
Formulae (\ref{angle-refr1}), (\ref{angle-refr2}), (\ref{angle-refr3}) are equivalent. In Appendix 2 we show how to transform formula (\ref{angle-refr1}) to (\ref{angle-refr2}) and (\ref{angle-refr3}).

In the formula (\ref{angle-refr2}) $N=N(r)$ and $r=\sqrt{b^2+z^2}$, so the expression under the integral sign is a function of $b$ and $z$. To calculate the deflection angle, we perform partial differentation with respect to $b$ at constant $z$ and then perform integration with respect to $z$ at constant $b$, and we obtain deflection angle as a function of $b$.

In the formula (\ref{angle-refr3}) we differentiate $N=N(r)$ with respect to $r$, then substitute $r=\sqrt{b^2+z^2}$. Expression under the integral sign will be a function of $b$ and $z$, and we perform integration along $z$ axis at constant $b$, and we obtain deflection angle as a function of $b$.

All observational predictions about the difference in angular positions and fluxes in different frequencies take place also in the case of a non-homogeneous plasma distribution, because refractive deflection in plasma is also chromatic. We should emphasize that the presence of homogeneous or non-homogeneous plasma increases the gravitational deflection of photon. Vacuum gravitational deflection is usually considered as positive ($\alpha_{einst}>0$) therefore the additional correction to gravitational deflection due to plasma presence is also positive ($\alpha_{add}>0$).
The refractive deflection can be both positive or negative, depending on the density profile. Usually the density of plasma in different models decreases with radius ($dN/dr<0$), therefore the refraction deflection is usually opposite to the gravitational deflection: the correction due to refractive deflection is negative ($\alpha_{refr}<0$), see \cite{BKTs2010}.

For example, in the case of inhomogeneous plasma with a power-behaved concentration
\begin{equation}
N(r) = N_0 (r_0/r)^k, \;  N_0 = \mbox{const}, \; r_0 = \mbox{const}, \; k=\mbox{const}>0
\end{equation}
the refractive deflection is (\citep{BKTs2009}, see also \cite{BliokhMinakov}, \cite{Muhleman1970}, \cite{Thompson}, \cite{zadachnik})
\begin{equation}
\alpha_{refr} = - \frac{K_e}{\omega^2} N_0 \left(\frac{R_0}{b}\right)^k \frac{\sqrt{\pi} \,
\Gamma\left(\frac{k}{2} + \frac{1}{2}\right)}{\Gamma\left(\frac{k}{2}\right)} \, ,
\end{equation}
\begin{equation}
\Gamma(x) = \int \limits_0^\infty t^{x-1} e^{-t} dt .
\end{equation}
Model with $k=1.25$ is one of two models used in paper Er and Mao \cite{Mao2014}.

In the case of axisymmetric mass distribution the deflection with the
impact parameter $b$ is equal to the Einstein angle for the mass
$M(b)$, where $M(b)$ is the projected mass enclosed by the circle of
the radius $b$ \cite{Clark, BliokhMinakov, GL, GL2}. In another words it is mass inside the cylinder with
the radius $b$. So, for this
case, we should write:
\begin{equation} \label{projected-mass}
\alpha_{einst} = \frac{4 M(b)}{b} \, \; \; \mbox{or, in usual units} \; \;\frac{4 G M(b)}{c^2 b} \, .
\end{equation}\\

\section{Gravitational lensing in plasma in the limit of strong deflection: analytic formulae}

Influence of plasma on the relativistic images have been considered for the first time in paper \cite{TsBK2013}, where
we have considered the case of a strong deflection angle for the light, traveling near the Schwarzschild black hole, surrounded by a uniform plasma. We derive for the first time the asymptotic analytic formulae for the gravitational deflection angle of photons in the homogeneous plasma, in a strong deflection limit.

We have obtained the deflection angle $\hat{\alpha}$ as a function of the closest approach distance $R$ and the ratio of frequencies
$\omega_e^2/\omega^2$ in the form \cite{TsBK2013}
\begin{equation} \label{alpha-R-E}
\hat{\alpha}(R,x) = -2 \sqrt{\frac{1+x}{2x}} \ln \left[z_1(x) \, \frac{R-r_M}{r_M}  \right] - \pi \, ,
\end{equation}
where
\begin{equation} \label{z-x}
\quad z_1(x) = \frac{9x-1+2\sqrt{6x(3x-1)}}{48x} \, , \;
r_M=6M \, \frac{1+x}{1+3x} \, , \; x = \sqrt{1-\frac{8 \omega_e^2}{9 \omega^2}} \, .
\end{equation}
Here $r_{M}$ is a critical (minimal) distance of the closest approach $R$, at given $\omega_e^2/\omega^2$. This formula (\ref{alpha-R-E}) is asymptotic and is valid for $R$ close to $r_M$. At $\omega \gg \omega_e$ the expression (\ref{alpha-R-E}) becomes the vacuum formula (\ref{vacuum-alpha-R}). At the given ratio of frequencies $\omega_e^2/\omega^2$, the trajectory is determined by a choice of $R$. The deflection angle of a photon goes to infinity when $R$ goes to $r_M$, and photon performs infinite number of turns at circle with the radius $r_{M}$. If $\omega \gg \omega_e$, then $x \to 1$, giving $r_{M} = 3M$, what corresponds to  photons in the vacuum. The presence of a homogeneous plasma increases the radius of the critical orbits of the photons around a black hole.

We have also obtained the deflection angle $\hat{\alpha}$ as a function of the impact parameter $b$ and the ratio of frequencies
$\omega_e^2/\omega^2$:
\begin{equation} \label{alpha-b-E}
\hat{\alpha}(b,x) = - \sqrt{\frac{1+x}{2x}} \, \ln \left[ \frac{2\,z_1^2(x)}{3x} \, \frac{b-b_{cr}}{b_{cr}} \right] - \pi \, , \quad
b_{cr} = \sqrt{3} \, r_M \, \sqrt{\frac{1+x}{3x-1}} \, .
\end{equation}
This formula is valid for $b$ close to $b_{cr}$, $b_{cr}$ is a critical value of impact parameter at given $\omega_e^2/\omega^2$. For $\omega \gg \omega_e$ we obtain the critical impact parameter for vacuum: $b_{cr} = 3 \sqrt{3} M$. At $\omega \gg \omega_e$ the expression (\ref{alpha-b-E}) becomes the vacuum formula (\ref{vacuum-alpha-b}).

\begin{figure}
\centerline{\hbox{\includegraphics[width=0.4\textwidth]{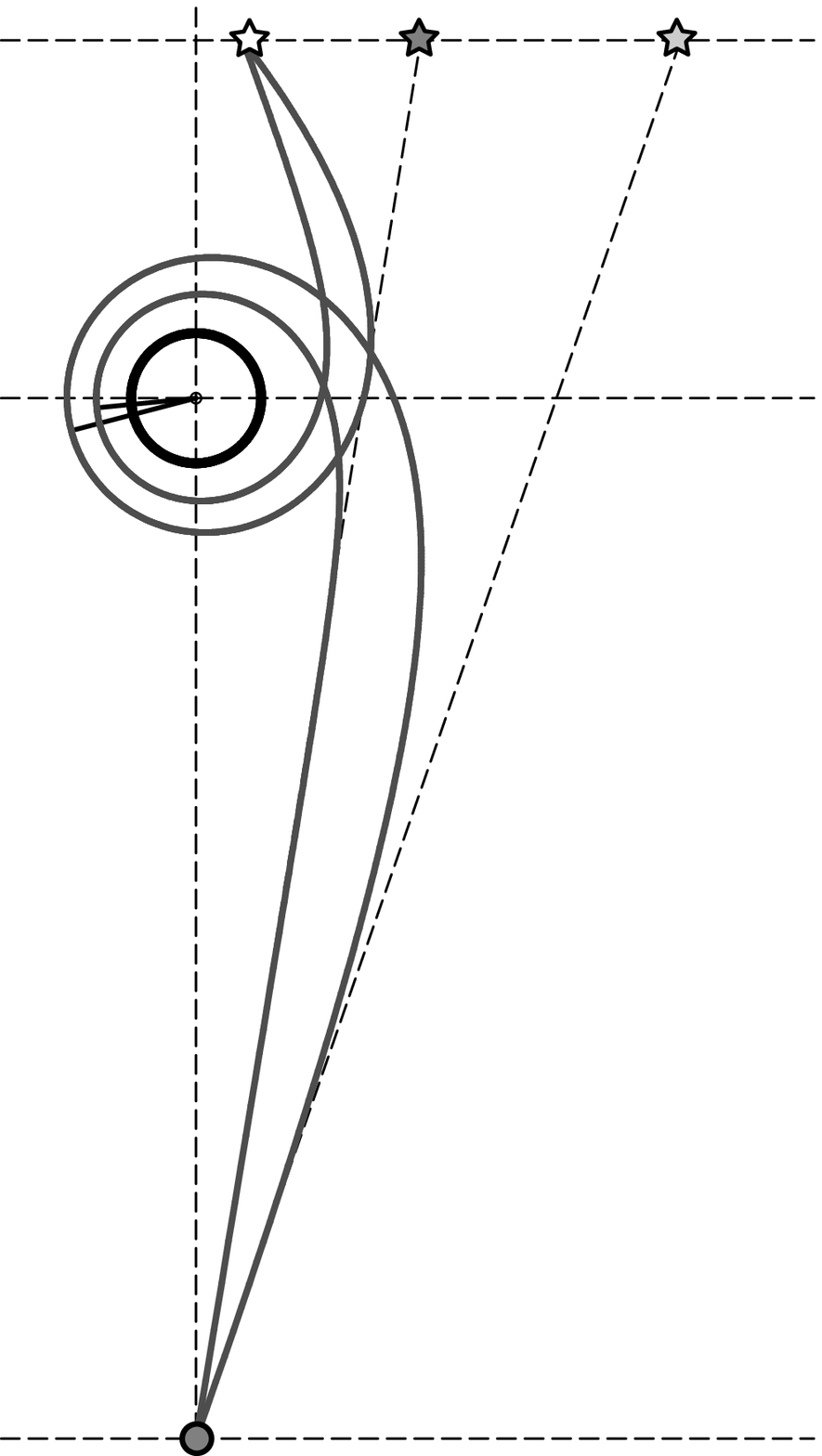}}} \caption{
Gravitational lensing in homogeneous plasma. Scheme
of formation of the relativistic images of different frequencies.
The photons of smaller frequency are deflected by a larger
angle by the gravitating center. The first ray has the frequency
$\omega \gg \omega_e$, in this case plasma effects are negligible, and the
trajectory can be computed using the vacuum equations
(see Fig.\ref{figure-prd2}, geometry and positions of objects are the same). The other ray has the frequency $\omega \simeq \omega_e$, in this case plasma effects are significant, and the trajectory
should be computed using plasma equations.} \label{figure-prd4}
\end{figure}

We apply these formulae for the calculation of positions and magnifications of relativistic images in the homogeneous plasma \cite{TsBK2013}, which are formed by the photons performing one or several revolutions around the central object (see Fig.\ref{figure-prd4}). For simplicity, let us rewrite formula (\ref{alpha-b-E}) as:
\begin{equation} \label{alpha-simple}
\hat{\alpha}(b,x) = - a(x) \, \ln \left( \frac{b-b_{cr}}{b_{cr}} \right) + c(x) \, ,
\end{equation}
where $a(x)$ and $c(x)$ are written as
\[
a(x) = \sqrt{\frac{1+x}{2x}} \, , \quad c(x) = - \sqrt{\frac{1+x}{2x}} \, \ln \left[ \frac{2\,z_1^2(x)}{3x} \right] - \pi \, ,
\]
$z_1(x)$, $r_M$ and $x$ are defined in (\ref{z-x}), $b_{cr}$ is defined in (\ref{alpha-b-E}).

To obtain the impact parameters corresponding to the relativistic images we can write an equation:
\begin{equation} \label{2-pi-n}
\hat{\alpha}(b,x) = 2 \pi n \, , \quad n=1,2, ...
\end{equation}
Here $n$ is number of pair of relativistic images. More rigorously, the deflection angle in this case is $\hat{\alpha} = 2 \pi + \Delta \hat{\alpha}$ (see Figures \ref{figure-prd2} and \ref{figure-prd4}), where $\Delta \hat{\alpha}$ has different values for different positions of source. But since
$\Delta \hat{\alpha} \ll 2 \pi n$ we can write equation for $b$ in the form (\ref{2-pi-n}), with high accuracy. We obtain the
impact parameters $b_n(x)$ of the relativistic images:
\begin{equation}
b_n = b_{cr} \left[ 1 + \exp \left(\frac{c(x)-2\pi n}{a(x)} \right) \right] \,  .
\end{equation}
The corresponding angular positions of the relativistic images are:
\begin{equation} \label{theta-n}
\theta_n = \frac{b_{cr}}{D_d} \left[ 1 + \exp \left(\frac{c(x)-2\pi n}{a(x)} \right) \right]  \,  ,
\end{equation}
where $D_d$ is the distance between observer and lens.

The angular positions $\theta_n$ in plasma is always bigger than the positions in vacuum. In Table 1 results for $E = 2$ ($\omega =2 \omega_e$) are presented. Therefore the presence of homogeneous plasma \textit{increases} the angular separation of the point relativistic images from gravitating center or the angular size of the relativistic rings.

Let us consider $b_n$ near the vacuum values. With $E^2 \rightarrow \infty$ ($\omega \gg  \omega_e$) we have:
\begin{equation}
r_M \simeq 3M \left( 1 + \frac{1}{9E^2}  \right) , \;\; b_{cr} \simeq 3 \sqrt{3} M \left(  1 + \frac{1}{3E^2} \right) , \;\; \theta_n = \frac{b_n}{D_d} \, ,
\end{equation}
\begin{equation}
b_n \simeq 3\sqrt{3} M (1+l_n^{vac})  \,
+ \, 3\sqrt{3} M \frac{1}{9E^2} \left\{ 3 + l_n^{vac} \left[ \pi(2n+1) + 2\sqrt{3} -3   \right] \right\} \, ,
\end{equation}
where
\begin{equation} \label{e-n-v}
l_n^{vac} = 216 (7 - 4\sqrt{3}) \, e^{-\pi (2n+1)} \, .
\end{equation}
Note that $E=\infty$ corresponds to the vacuum.

Let us consider a behavior of $b_n$ in opposite situation: when the photon frequency is close to plasma frequency. With $E^2 \rightarrow 1$ ($\omega \simeq  \omega_e$) we have:
\begin{equation}
r_M \simeq 4M [ 1 + (E^2-1)  ]   \, , \; \;
b_{cr} \simeq \frac{4M}{\sqrt{E^2-1}} \simeq \frac{4M}{n_{pl}}  \; \left( \mbox{here} \; \; n_{pl}=\sqrt{1-\frac{\omega_e^2}{\omega^2}} \right) ,
\end{equation}
\begin{equation}
\theta_n = \frac{b_n}{D_d} \, , \; b_n \simeq \frac{4M}{\sqrt{E^2-1}}  \left[  1+  32 \, e^{-\pi (2n+1)/ \sqrt{2}} \right] .
\end{equation}
We see that with $\omega \rightarrow  \omega_e$ angular sizes $\theta_n$ increases unboundedly.

\begin{table}[h]
\caption{\label{table1}
Comparison of the impact parameters and the angular sizes of the relativistic rings in vacuum and homogeneous plasma for $\omega = 2 \omega_e$ ($E =2$).  For convenience, differences $b_n-b_{cr}$ are presented. The values of impact parameters are given in units of mass $M$. The angular sizes can be calculated as $\theta = b/D_d$. These values can also be considered as positions of two relativistic images of the point source, in case if there is no perfect alignment.}
\begin{ruledtabular}
\begin{tabular}{lcc}
Relativistic rings & vacuum & plasma \\
\colrule
Critical value ($b_{cr}$) & $3\sqrt{3}$ & $3\sqrt{3} + 0.540300928156$ \\
Ring 1 ($b_1-b_{cr}$) & 0.00650 & $0.00899$ \\
Ring 2 ($b_2-b_{cr}$) & $1.21 \cdot 10^{-5}$ & $2.05 \cdot 10^{-5}$ \\
Ring 3 ($b_3-b_{cr}$) & $2.27 \cdot 10^{-8}$ & $4.68 \cdot 10^{-8}$ \\
\end{tabular}
\end{ruledtabular}
\end{table}

Let us now consider the magnification of the relativistic images. We consider the relativistic images of a point source located at the angular position $\beta$ from line connecting the observer and the gravitational center (lens). We use approximate approach, similar to \cite{Bozza2002}, based on our analytical formula (\ref{alpha-simple}), see \cite{TsBK2013} for details. More precise way is to use the exact formula (\ref{perlick-angle}) and to solve this problem by the similar approach as it was done in \cite{Virbhadra2000} for relativistic images in vacuum.

We obtain:
\begin{equation} \label{mu-n}
\mu_n = \frac{D_s b_{cr}^2 (1+ l_n) l_n}{D_{ds} D_d^2  \, a(x) \, \beta} \, , \quad
l_n = \exp \left(\frac{c(x)-2\pi n}{a(x)} \right) ,
\end{equation}
and $a(x)$ and $c(x)$ are coefficients defined in (\ref{alpha-simple}). In expression (\ref{mu-n}) the variables $b_{cr}$, $l_n$, $a(x)$, $c(x)$ depend on $x$, so these variables depend on the ratio of the photon and the plasma frequencies. Magnification $\mu_n$ of the relativistic images of the point source formally tends to infinity if the source angular position $\beta$ goes to 0, as it takes place in vacuum \cite{Virbhadra2000}.

In case of vacuum ($\omega/\omega_e=\infty$):
\begin{equation} \label{mu-n-v}
\mu_n^{vac} = \frac{D_s b_{cr}^2 (1+ l_n^{vac}) l_n^{vac}}{D_{ds} D_d^2  \, \beta} \, , \quad b_{cr} = 3\sqrt{3} M \, ,
\end{equation}
where $l_n^{vac}$ is defined in (\ref{e-n-v}).

If $E^2 \rightarrow 1$  ($\omega \simeq \omega_e$), we have:
\begin{equation}
\mu_n = \frac{D_s}{D_{ds} D_d^2} \, \frac{16M^2}{E^2-1}   \,  \frac{(1+l_n^{pl})  l_n^{pl}}{\sqrt{2} \, \beta} \, , \quad
l_n^{pl} = 32 e^{-\pi(2n+1)/\sqrt{2}} \, .
\end{equation}
We see that if $\omega \rightarrow \omega_e$, magnifications $\mu_n$ can increase unboundedly.

Numerical examples are given in Table 2.

Possibility and difficulties of observation of relativistic images in vacuum are discussed in big details in \cite{Virbhadra2000} and \cite{Virbhadra2009}. Presence of plasma  formally may lead to much stronger magnification of relativistic images at $\omega \rightarrow \omega_e$, but increasing absorbtion creates additional problems for observations.

While there are many observations of multiple 'ordinary' \, images formed by photons weak deflected by massive lenses, the observations of relativistic images formed by the photons in strong deflection regime seems to be very unlikely. This low probability becomes even less, if we take into account that there are reliable observations of black holes only with accretion disc environment which significantly absorbes the radiation.

Nevertheless, existence of isolated black holes in edge of Galaxy can not be fully excluded. Detection of them by usual methods is almost impossible, but lensing of the distant source radiation (for example, star of Magellan Clouds) by such black hole might be observed with using some non-trivial technique. This detection is unlikely, but if it occurs, it would be very informative both for finding the position of lensing object and for properties of the surrounding medium.

\begin{table}[h]
\caption{\label{table1}
Comparison of the magnification factors of relativistic images for lensing in homogeneous plasma and in vacuum. Ratios $\mu_n/\mu_n^{vac}$ are presented, at given $D_s$, $D_d$, $D_{ds}$, $\beta$. Values $\mu_n$ and $\mu_n^{vac}$ are calculated with (\ref{mu-n}) and (\ref{mu-n-v}) correspondingly. Values $\mu_n$ and $\mu_n^{vac}$ depend on $D_s$, $D_d$, $D_{ds}$, $\beta$, $\omega/\omega_e$, but ratio $\mu_n/\mu_n^{vac}$ depends only on $\omega/\omega_e$. In last column the values $\mu_n^{vac}$ are calculated for $M/D_d = 2.26467 \cdot 10^{-11}$, what corresponds to supermassive black hole in center of Milky Way, $D_s/D_{ds}=2$, $\beta = 1 \mu$as (these parameters have been taken from \cite{Virbhadra2009}).    }
\begin{ruledtabular}
\begin{tabular}{lcccc}
$\mu_n / \mu_n^{vac}$ &
$\omega/\omega_e=1.1$ &
$\omega/\omega_e=2$   &
$\omega/\omega_e=10$  &
$\mu_n^{vac}$ \\
\colrule
$\mu_1 / \mu_1^{vac}$  &  $13.3$ & 1.48 & 1.01 & $ 0.716 \cdot 10^{-11}$ \\
$\mu_2 / \mu_2^{vac}$  &  $40.0$ & 1.81 & 1.02 & $ 0.134 \cdot 10^{-13}$ \\
$\mu_3 / \mu_3^{vac}$  &  $120$  & 2.21 & 1.03 & $ 0.249 \cdot 10^{-16}$ \\
\end{tabular}
\end{ruledtabular}
\end{table}

\section{Conclusions}

(i) The model of gravitational lensing in plasma is developed in details, by using Synge's hamiltonian theory for geometrical optic in gravity and medium. Exact formula for the deflection angle in the Schwarzschild metric, surrounded by plasma with spherically symmetric density distribution is written as (\ref{perlick-angle}), without the limit for the value of the deflection angle, the distance of the closest approach, or the ratio of plasma and photon frequencies. In the case of  a homogeneous plasma the integral for the deflection angle is expressed via elliptic integrals, see Section 5.

(ii) In presence of both gravity and plasma the deflection angle is physically defined by mutual combination of different phenomena: gravity, dispersion, refraction. Effects of deflection by gravity in vacuum and the refractive deflection in non-homogeneous medium are well known and presented in literature with necessary details. New effect, which we emphasize, is that in the case of a homogeneous plasma, in absence of refractive deflection, the gravitational deflection differs from the vacuum deflection, see (\ref{angle-plasma-R}) and (\ref{angle-plasma-b}).

(iii) Plasma is a dispersive medium, its refractive index depends on the photon frequency. Therefore the deflection angle in presence of both gravity and plasma always depends on the photon frequency. In particular, presence of homogeneous plasma leads to chromatic gravitational deflection; the presence of non-homogeneous plasma leads also to chromatic gravitational deflection and to chromatic refractive deflection. So \textit{presence of plasma always makes gravitational lensing chromatic and leads to difference in angular position of the same image at different wavelengths}.

(iv) For applications to gravitational lensing problems connected with observations, see section \ref{section-uniform} and \ref{section-nonuniform}  where case of weak deflection is discussed. For the deflection by point-mass in a homogeneous plasma see (\ref{angle-plasma-b}), for the inhomogeneous case  see (\ref{angle-nonhomogen}) with (\ref{einst-b}), (\ref{angle-refr1}), (\ref{angle-refr2}), (\ref{angle-refr3}). In the case of gravitating mass distribution (in our formulae we neglect the mass of plasma particles) the projected mass $M(b)$ should be used (\ref{projected-mass}), as it is usually done in lensing. As an example of application of these formulae, see numerical modelling of gravitational lens with plasma which is performed in \cite{Mao2014}, with attention to many gravitational lens phenomena and discussion of observational possibilities.

(v) The deflection angles for light rays near Schwarzschild black hole in homogeneous plasma, in strong deflection limit, are derived. Using of strong deflection limit allows us to investigate analytically properties of relativistic images in presence of plasma.

\section*{Acknowledgments}

Authors are grateful to A.V. Byalko for useful comments.

The work of GSBK and OYuT was partially supported by the Russian Foundation for Basic Research Grant No. 14-02-00728 and the Russian Federation President Grant for Support of Leading Scientific Schools, Grant No. NSh-261.2014.2.

The work of OYuT was also partially supported by the Russian Federation President Grant for Support of Young Scientists, Grant No. MK-2918.2013.2.

\appendix

\section{Appendix 1}
Let us rewrite formula (\ref{perlick-angle}) as
\begin{equation}
\hat{\alpha} = 2 \int \limits_R^\infty \frac{dr}{r\sqrt{1-\frac{2M}{r}}\sqrt{\frac{h^2(r)}{h^2(R)} - 1 }} - \pi \, ,
\end{equation}
with
\begin{equation}
h(r)  = r \sqrt{ \frac{1}{1-2M/r} - \frac{\omega_e^2(r)}{\omega^2} } \, .
\end{equation}

We consider case with $\omega_e^2= \mbox{const}$. We denote: $\omega_e^2/\omega^2=B$. Using that $R \gg M$ and $r \gg M$, we obtain:
\begin{equation}
\frac{h^2(r)}{h^2(R)} \simeq \frac{r^2 \left( 1 + \frac{2M}{r} - B \right)}{R^2 \left( 1 + \frac{2M}{R} - B  \right)} = \frac{r^2 \left( 1 + \frac{2M}{r(1-B)}  \right) }{R^2 \left( 1 + \frac{2M}{R(1-B)}  \right)} \simeq
\end{equation}
\[
\simeq \frac{r^2}{R^2} \left( 1 + \frac{2M}{r(1-B)}   - \frac{2M}{R(1-B)}  \right) = \frac{r^2}{R^2} \, + \,  \frac{r^2}{R^2} \, \frac{2M}{1-B} \, \frac{R-r}{rR}
\]
Transforming
\begin{equation}
\left[ \frac{h^2(r)}{h^2(R)} - 1 \right]^{-1/2} = \left[  \frac{r^2-R^2}{R^2} + \frac{2M}{1-B} \, \frac{r(R-r)}{R^3}    \right]^{-1/2}   \simeq
\end{equation}
\[
\simeq   \frac{R}{\sqrt{r^2-R^2}} \left[  1 + \frac{M}{1-B} \, \frac{r(r-R)}{R(r^2-R^2)} \right]
\]
and
\begin{equation}
\frac{1}{r \sqrt{1-2M/r}}  \simeq  \frac{1}{r}  \left( 1 + \frac{M}{r}  \right) ,
\end{equation}
we get
\begin{equation}
\hat{\alpha} = 2 \int \limits_R^\infty  \frac{R}{r \sqrt{r^2-R^2}} \, dr  +  2 \int \limits_R^\infty  \frac{R}{r \sqrt{r^2-R^2}}  \frac{M}{r} \, dr  +  2 \int \limits_R^\infty   \frac{M}{1-B} \frac{r-R}{(r^2-R^2)^{3/2}} \, dr - \pi  =
\end{equation}
\[
= \pi + \frac{2M}{R}  + 2 \frac{M}{1-B} \, \frac{1}{R}   -  \pi = \frac{2M}{R} \left( 1 + \frac{1}{1 - \omega_e^2/\omega^2} \right) .
\]

\section{Appendix 2}
Let us expand formula (\ref{perlick-angle}) with $M/r \ll 1$ and $\omega_e^2(r)/\omega^2 \ll 1$. We denote $\omega_e^2(r)/\omega^2 = B_r$ and $\omega_e^2(R)/\omega^2 = B_R$. We get:
\begin{equation} \label{angle-app2}
\hat{\alpha} = \alpha_{straight}  +   \alpha_{einst} +  \alpha_{refr} - \pi =   2 \int \limits_R^\infty  f_0 \, dr + 2 \int \limits_R^\infty  f_1 \, dr + 2 \int \limits_R^\infty  f_2 \, dr  -  \pi \, ,
\end{equation}
where
\begin{equation}
f_0 = \frac{R}{r\sqrt{r^2-R^2}} ,
\end{equation}
\begin{equation}
f_1 = \frac{(r^2 + rR + R^2)M}{r^2(r+R)\sqrt{r^2-R^2}} ,
\end{equation}
\begin{equation}
f_2 = \frac{rR}{2(r^2-R^2)^{3/2}} (B_r-B_R) .
\end{equation}
Integration of $f_0$ leads to $\alpha_{straight} = \pi$ which is cancelled by $\pi$ in (\ref{angle-app2}). Integration of $f_1$ leads to $\alpha_{einst} = 2M/R$ which can be written here as $2M/b$. Integration of $f_2$ is:
\begin{equation}
\alpha_{refr} =  \int \limits_R^\infty  \frac{rR}{(r^2-R^2)^{3/2}} (B_r-B_R) \, dr  = -R \int \limits_R^\infty  (B_r-B_R) \, d \left( \frac{1}{\sqrt{r^2-R^2}} \right)  =
\end{equation}
\[
= - \left. \frac{R(B_r-B_R)}{\sqrt{r^2-R^2}}  \right|_R^\infty + R \int \limits_R^\infty \frac{1}{\sqrt{r^2-R^2}} \frac{dB_r}{dr} \, dr \, .
\]
The first term is evidently equals to zero at $r=\infty$. At $r \rightarrow R$ we can write:
\begin{equation}
\left. \frac{R(B_r-B_R)}{\sqrt{r^2-R^2}} \right|_{r \rightarrow R}   \simeq
\left.  \frac{1}{\sqrt{2R(r-R)}}  \, R \left( B_R + \left. \frac{dB_r}{dr} \right|_{r=R}  (r-R)  - B_R   \right) \right|_{r \rightarrow R} \propto \left. \sqrt{r-R}  \right|_{r \rightarrow R}  = 0 ,
\end{equation}
and we obtain:
\begin{equation}
\alpha_{refr} = R \int \limits_R^\infty \frac{1}{\sqrt{r^2-R^2}} \frac{dB_r}{dr} \, dr  =   \frac{R K_e}{\omega^2} \int \limits_R^\infty \frac{1}{\sqrt{r^2-R^2}} \frac{dN(r)}{dr} \, dr .
\end{equation}

Let us transform this formula to variables $b$ and $z$. We consider an unperturbed trajectory as straight line parallel axis $z$ with impact parameter $b$, and $r=\sqrt{b^2 + z^2}$. We transform $r$-derivative to derivative with respect to perpendicular to $z$ direction which we can denote as $b$:
\begin{equation}
\frac{\partial N}{\partial b} = \frac{dN}{dr} \frac{\partial r}{\partial b} =   \frac{dN}{dr} \frac{b}{r}
\end{equation}
We transform integration with respect to $r$ to integration with respect to $z$, at constant $b$:
\begin{equation}
dr = d (\sqrt{b^2+z^2}) = \frac{z \, dz}{r}
\end{equation}
We can change $R$ to $b$.

Substituting
\begin{equation}
R=b, \; \sqrt{r^2-R^2} = z, \; \frac{dN}{dr} = \frac{r}{b} \frac{\partial N}{\partial b} , \; dr = \frac{z \, dz}{r} ,
\end{equation}
we obtain
\begin{equation}
\alpha_{refr} = \frac{b K_e}{\omega^2}   \int \limits_0^\infty \frac{1}{z} \frac{r}{b} \frac{\partial N}{\partial b} \frac{z}{r} \, dz = \frac{K_e}{\omega^2}   \int \limits_0^\infty \frac{\partial N}{\partial b} \, dz
\end{equation}
and
\begin{equation}
\alpha_{refr} = \frac{K_e b}{\omega^2}   \int \limits_0^\infty \frac{1}{r} \frac{dN}{dr} \, dz .
\end{equation}


\begin{thebibliography}{99}



\bibitem{Walsh1979}
D. Walsh, R. F. Carswell and R. J. Weymann, Nature, \textbf{279}, 381 (1979).

\bibitem{Castles}
C.S. Kochanek, E.E. Falco, C. Impey, J. Lehar, B. McLeod, H.-W. Rix, CASTLES Survey, http://www.cfa.harvard.edu/castles/

\bibitem{BKTs2009}
G. S. Bisnovatyi-Kogan and O. Yu. Tsupko, Gravitation and Cosmology, \textbf{15}, 20 (2009).

\bibitem{BKTs2010}
G. S. Bisnovatyi-Kogan and O. Yu. Tsupko, Mon. Not. R. Astron. Soc. \textbf{404}, 1790 (2010).

\bibitem{TsBK2012}
O. Yu. Tsupko and G. S. Bisnovatyi-Kogan, Gravitation and Cosmology, \textbf{18}, 117 (2012).

\bibitem{TsBK2013}
O. Yu. Tsupko and G. S. Bisnovatyi-Kogan, Physical Review D \textbf{87}, 124009 (2013).



\bibitem{Kayser1986}
R. Kayser, S. Refsdal, and R. Stabell, Astronomy and Astrophysics, \textbf{166}, 36 (1986)

\bibitem{WambsganssPaczynski1991}
J. Wambsganss and B. Paczy\'{n}ski, Astronomical Journal, \textbf{102}, 864 (1991)



\bibitem{Muhleman1970}
D. O. Muhleman, R. D. Ekers, and E. B. Fomalont, Phys. Rev. Lett. \textbf{24}, 24, 1377 (1970).

\bibitem{Muhleman1966}
D. O. Muhleman and I. D. Johnston, Phys. Rev. Lett. \textbf{17}, 8, 455 (1966).

\bibitem{zadachnik}
A. P. Lightman, W. H. Press, R. H. Price, and S. A. Teukolsky, Problem Book in Relativity and Gravitation. Princeton University Press, Princeton, New Jersey, New York, 1979.

\bibitem{BliokhMinakov}
P. V. Bliokh and A. A. Minakov, Gravitational Lenses. Naukova Dumka, Kiev, 1989. (in Russian)


\bibitem{Synge}
J. L. Synge, Relativity: the General Theory. North-Holland Publishing Company, Amsterdam, 1960.


\bibitem{Bicak}
J. Bi\v{c}\'{a}k and P. Hadrava, Astron. and Astrophys. \textbf{44}, 389 (1975).

\bibitem{Krikorian1985}
S. Kichenassamy and R. A. Krikorian, Physical Review D \textbf{32}, 1866 (1985).

\bibitem{Krikorian1999}
R. A. Krikorian, Astrophysics \textbf{42} (3), 338 (1999)


\bibitem{Perlick2000}
V. Perlick, Ray Optics, Fermats Principle, and Applications to General Relativity. Springer-Verlag, Berlin, 2000.


\bibitem{Mao2014}
Xinzhong Er and Shude Mao, Monthly Notices of the Royal Astronomical Society, \textbf{437}, 2180 (2014).


\bibitem{morozova}
V. S. Morozova, B. J. Ahmedov, and A. A. Tursunov, Astrophysics and Space Science, \textbf{346}, 513 (2013).

\bibitem{Virbhadra2000}
K. S. Virbhadra and G. F. R. Ellis, Phys. Rev. D \textbf{62}, 084003 (2000).









\bibitem{GL}
P. Schneider, J. Ehlers, and E. E. Falco, Gravitational Lenses. Springer-Verlag, Berlin, 1992.

\bibitem{GL2}
P. Schneider, C. S. Kochanek, and J. Wambsganss, Gravitational Lensing: Strong, Weak and Micro, Swiss Society for Astrophysics and Astronomy Series: Saas-Fee Advanced Courses, Number 33. Springer, Berlin, 2006.

\bibitem{Perlick2004review}
V. Perlick, Living Rev. Relativ. \textbf{7}, 9 (2004)

\bibitem{Perlick2010}
V. Perlick, eprint arXiv:1010.3416 (2010)

\bibitem{BartelmannSchneider-review}
M. Bartelmann, P. Schneider, Physics Reports, \textbf{340}, 291 (2001)


\bibitem{Bartelmann-review}	
M. Bartelmann, Classical and Quantum Gravity, \textbf{27}, Issue 23, id. 233001 (2010).


\bibitem{Wambsganss-review}
J. Wambsganss, Living Reviews in Relativity, \textbf{1}, 12 (1998)


\bibitem{Hoekstra-review}	
H. Hoekstra, Bh. Jain, Annual Review of Nuclear and Particle Systems, \textbf{58}, Issue 1, 99 (2008)



\bibitem{Dyson1920}
F. W. Dyson, A. S. Eddington and C. Davidson, Philosophical Transactions of the
Royal Society of London. Series A, \textbf{220}, 291 (1920).


\bibitem{Refsdal1964b}	
S. Refsdal, Monthly Notices of the Royal Astronomical Society, \textbf{128}, 307 (1964)

\bibitem{Clowe2006}
D. Clowe \emph{et al.}, The Astrophysical Journal, \textbf{648}, Issue 2, L109-L113 (2006)


\bibitem{Refsdal1964a}	
S. Refsdal, Monthly Notices of the Royal Astronomical Society, \textbf{128}, 295 (1964)

\bibitem{Byalko1969}
A. V. Byalko, Astron. Zh. \textbf{46}, 998 (1969), (English translation: Sov. Astron. \textbf{13}, 784, (1970))



\bibitem{Mao1991}
S. Mao, B. Paczy\'{n}ski, ApJ, \textbf{374}, L37 (1991).


\bibitem{Beaulieu2006}
J.-P. Beaulieu, D. P. Bennett, P. Fouqu\'{e} \emph{et al}, Nature \textbf{439}, 437 (2006).








\bibitem{MTW}
C. W. Misner, K. S. Thorne, and J. A. Wheeler, Gravitation. Freeman, New York, 1973.

\bibitem{Darwin1959}
C. Darwin, Proceedings of the Royal Society of London, Series A, Mathematical and Physical Sciences, \textbf{249} (1257), 180 (1959)

\bibitem{Keeton}
C. R. Keeton and A. O. Petters, Phys. Rev. D \textbf{72}, 104006 (2005)

\bibitem{LL2}
L. D. Landau and E. M. Lifshitz , The Classical Theory of Fields. Pergamon, Oxford, 1993.

\bibitem{Bozza2001}
V. Bozza, S. Capozziello, G. Iovane, and G. Scarpetta, Gen. Relativ. Gravit. \textbf{33}, 1535 (2001).

\bibitem{Bozza2002}
V. Bozza, Phys. Rev. D \textbf{66}, 103001 (2002).

\bibitem{Tsupko2014}
O. Yu. Tsupko, Physical Review D \textbf{89}, 084075 (2014)

\bibitem{BKTs2008}
G. S. Bisnovatyi-Kogan and O. Yu. Tsupko, Astrophysics, \textbf{51}, 99 (2008).

\bibitem{Frittelli2000}
S. Frittelli, T. P. Kling, and E. T. Newman, Phys. Rev. D \textbf{61}, 064021 (2000).

\bibitem{Perlick2004a}
V. Perlick, Phys. Rev. D \textbf{69}, 064017 (2004)

\bibitem{Bozza2008}
V. Bozza, Phys. Rev. D \textbf{78}, 103005 (2008).

\bibitem{PerlickMG}
V. Perlick, in Proceedings of the MG11 Meeting on General Relativity, edited by Hagen Kleinert, Robert T. Jantzen, editor of the Marcel Grossmann Meeting series: Remo Ruffini, published by World Scientific Publishing Co. Pte. Ltd., 2008, pp. 680-699, eprint arXiv:0708.0178 (2007)

\bibitem{Perlick2002}
W. Hasse and V. Perlick, Gen Relativ Gravit \textbf{34}, 415 (2002)

\bibitem{Virbhadra2009}
K. S. Virbhadra, Phys. Rev. D \textbf{79}, 083004 (2009).

\bibitem{Virbhadra2001}
Clarissa-Marie Claudel, K. S. Virbhadra, and G. F. R. Ellis, Journal of Mathematical Physics \textbf{42}, 818 (2001).

\bibitem{Virbhadra2008}
K. S. Virbhadra, C. R. Keeton, Phys. Rev. D \textbf{77}, 124014 (2008).

\bibitem{Bozza2010}
V. Bozza, Gen. Relativ. Gravit. \textbf{42}, 2269 (2010).

\bibitem{Eiroa2002}
E. F. Eiroa, G. E. Romero, and D. F.Torres, Physical Review D \textbf{66}, 024010 (2002)

\bibitem{IyerPetters}
S. V. Iyer and A. O. Petters, Gen Relativ Gravit \textbf{39}, 1563 (2007)

\bibitem{EiroaSendra}
E. F. Eiroa and C. M Sendra, Class. Quantum Grav. \textbf{28}, 085008 (2011)

\bibitem{Cite1}
N. Mukherjee and A. S. Majumdar, Gravitation and Cosmology, \textbf{15} (3), 263 (2009)

\bibitem{Cite2}
T. Ghosh and S. SenGupta, Physical Review D, \textbf{81} (4), 044013 (2010)

\bibitem{Cite3}
Shao-Wen Wei, Yu-Xiao Liu, Chun-E Fu, and Ke Yang, eprint arXiv: 1104.0776 (2011)





\bibitem{Kulsrud-Loeb}
R. Kulsrud and A. Loeb, Physical Review D, \textbf{45}, 525 (1992)

\bibitem{Brod-Blandford1}
A. Broderick and R. Blandford,  MNRAS, \textbf{342}, 1280 (2003)

\bibitem{Brod-Blandford2}
A. Broderick and R. Blandford, Ap\&SS, \textbf{288}, 161 (2003)




\bibitem{Weinberg}
S. Weinberg, Gravitation and cosmology: principles and applications of the general theory of relativity. John Wiley and Sons, Inc., New York -- London -- Sydney -- Toronto, 1972.

\bibitem{Zeld-Novikov}
Ya. B. Zel'dovich and I. D. Novikov, The Theory of Gravitation and Stellar Evolution. Nauka, Moscow, 1971. (in Russian)


\bibitem{Chandra}
S. Chandrasekhar, The mathematical theory of black holes. Clarendon Press Oxford, Oxford University Press, New York, 1983.





\bibitem{Hagihara1931}
Yu. Hagihara, Japanese Journal of Astronomy and Geophysics, \textbf{8}, 67 (1931)

\bibitem{Metzner1963}
A. W. K. Metzner, Journal of Mathematical Physics, \textbf{4} (9), 1194 (1963)

\bibitem{M-Pleb1962}
B. Mielnik and J. Pleba\'{n}ski, Acta Phys. Polon. \textbf{21}, 239 (1962)

\bibitem{Darwin1961}
C. Darwin, Proceedings of the Royal Society of London, Series A, Mathematical and Physical Sciences, \textbf{263} (1312), 39 (1961)

\bibitem{Bogorodsky1962}
A. F. Bogorodsky, Einstein's Field Equations and Their Application to Astronomy. Kiev Univ., Kiev, 1962. (in Russian)

\bibitem{MiroRodriguez1}
C. Miro Rodr\'{\i}guez, Nuovo Cimento B, \textbf{98} (1), 87 (1987)

\bibitem{MiroRodriguez2}
C. Miro Rodr\'{\i}guez, Nuovo Cimento B, \textbf{100} (6), 801 (1987)

\bibitem{Bronstein}
I. M. Bronstein and K. A. Semendyaev, Handbook of Mathematics. Van Nostrand, New York, 1985.

\bibitem{LLmech}
L. D. Landau and E. M. Lifshitz, Mechanics. Nauka, Moscow, 1988. (in Russian)

\bibitem{Korn}
G. A. Korn and T. M. Korn, Mathematical handbook for scientists and engineers: definitions, theorems, and formulas for reference and review. Courier Dover Publications, 2000.

\bibitem{Gr-Ryzhik}
I. S. Gradshteyn and I. M. Ryzhik, Table of integrals, series and products. Academic Press, New York, 1965.



\bibitem{Thompson}
C. Thompson, R. D. Blandford, Ch. R. Evans, and E. S. Phinney, ApJ, \textbf{422}, 304, (1994), Appendix A

\bibitem{Clark}
E. E. Clark, MNRAS, \textbf{158}, 233 (1972)






\end{thebibliography}
\end{document}